\documentclass[a4paper,fleqn,usenatbib]{mnras}

\usepackage[T1]{fontenc}
\usepackage{ae,aecompl}

\usepackage{graphicx}	
\usepackage{amsmath}	
\usepackage{amssymb}	

\usepackage{subfig}
\usepackage{url}
\usepackage{textcomp}
\usepackage{siunitx}
\usepackage{lscape}
\usepackage{hyperref}

\newcommand{\marc}{mag~arcsec$^{-2}$~}

\newcommand{\HI}{\textsc{HI}}
\newcommand{\molH}{\textsc{H}$_2$} 
\newcommand{\ttlgas}{$\Sigma_{\HI+\textsc{H}_2}$}
\newcommand{\sigmolH}{$\Sigma_{\textsc{H}_2}$}
\newcommand{\sighi}{$\Sigma_{\HI}$}
\newcommand{\CO}{\textsc{CO}}

\newcommand{\kms}{\mathrm{km\ s^{-1}}}
\newcommand{\mmu}{\mathrm{\mu m}}

\newcommand{\sigsfr}{$\Sigma_{SFR}$}

\newcommand{\mpc}{$\mathrm{M_\odot\ pc^{-2}}$}
\newcommand{\mhi}{$\mathrm{M_{\HI}}$}

\title[Scaling Relations in WISE-WHISP Galaxies]{\HI\ Global Scaling Relations in the WISE-WHISP Survey}

\author[Naluminsa et al.]{
E. Naluminsa,$^{1}$\thanks{E-mail: elizabeth.naluminsa@uct.ac.za (EN)}
E. C. Elson,$^{2}$
T. H. Jarrett$^{1}$
\\
$^{1}$Department of Astronomy, University of Cape Town, Private Bag X3, Rondebosch 7701, South Africa\\
$^{2}$Department of Physics \& Astronomy, University of the Western Cape, Robert Sobukwe Rd, Bellville 7535, South Africa\\
}

\date{Accepted 2021 January 04 . Received 2020 December 24; in original form 2020 July 12.}

\pubyear{2021}
\begin{document}
\label{firstpage}
\pagerange{\pageref{firstpage}--\pageref{lastpage}}
\maketitle

\begin{abstract}
We present the global scaling relations between the neutral atomic hydrogen gas, the stellar disk and the star forming disk in a sample of 228 nearby galaxies that are both spatially and spectrally resolved in \HI\ line emission. We have used \HI\ data from the Westerbork survey of \HI\ in Irregular and Spiral galaxies (\textsc{WHISP}) and Mid Infrared (3.4 $\mmu$, 11.6 $\mmu$) data from the Wide-field Infrared Survey Explorer (WISE) survey, combining two datasets that are well-suited to such a study in terms of uniformity, resolution and sensitivity. We utilize a novel method of deriving scaling relations for quantities enclosed within the stellar disk rather than integrating over the \HI\ disk and find the global scaling relations to be tighter when defined for enclosed quantities. We also present new \HI\ intensity maps for the WHISP survey derived using a robust noise rejection technique along with corresponding velocity fields.
\end{abstract}




\section{Introduction}
The fundamental goal of astrophysical studies is a better understanding of the origins and evolution of our universe. Both theoretical and observational works are geared towards constructing a solid picture of the processes involved in the formation of cosmic structures. This involves pursuing accurate models of galaxy formation and evolution. 
Central to the evolution of galaxies is the process of star formation (SF) which must be properly accounted for in models of both chemical and physical evolution (e.g. \citealt{1999MNRAS.307..857B,2002MNRAS.330..821L,2017MNRAS.467..115D}). This is because SF drives the consumption of gas in galaxies and the chemical and physical evolution of both the interstellar medium (ISM) and intergalactic medium (IGM) \citep{2015aska.confE.129D}. 

The scaling of SF (measured in terms of star formation rate, SFR,  and star formation efficiency, SFE) with other galaxy properties such as mass and mass surface density provides insights into how gas is utilized at difference epochs via collapse, accretion, ejection and recycling. A fundamental scaling relation is the Kennicutt-Schmidt (KS) law \citep{1959ApJ...129..243S,1989ApJ...344..685K,1998ApJ...498..541K} which relates the surface densities of SF and gas (atomic or molecular hydrogen) by a power law; 
\begin{center}
\begin{equation}
\mathrm{\Sigma_{SFR}\ =\ A \Sigma_g^N},
\label{eqn:kslaw}
\end{equation}
\end{center}
where $\mathrm{\Sigma_{SFR}}$ is the star formation rate surface density (in $\mathrm{M_\odot\ kpc^{-2}\ yr^{-1}}$), $\mathrm{\Sigma_g}$ is the gas surface density (in \mpc) and A is a constant of proportionality. 
\citet{1998ApJ...498..541K} found index N = 1.4$\pm$0.05 using a combination of \sighi\ and \sigmolH\ for total gas surface density. However, over the years, different studies have found varying values of N. Most notably, it has been shown that the value of N is closer to 1.0 when the surface density of molecular hydrogen gas (as traced by \CO ) is used instead of the total gas surface density, especially in molecule-rich ISM conditions such as the central regions of star forming spiral galaxies (e.g. \citealt{2002ApJ...569..157W,2008AJ....136.2846B,2013AJ....146...19L,2014A&A...566A.147D}). 
Studies such as \citet{2003MNRAS.346.1215B,2008AJ....136.2846B,2011AJ....142...37S,2018ApJ...852..106C} have also shown that there is, at best, a nonlinear power law relationship between \ttlgas\ and \sigsfr\ with a higher index N $\approx$ 2.0. 
This is attributed to the weak relationship between \sighi\ and \sigsfr\ because stars form from collapsing clouds of molecular hydrogen. Still, the atomic gas does have a connection to the SFR and it has been shown that reduction in the amount of \HI\ in galactic disks will suppress the SF \citep{1999AJ....118..670R,2002ApJ...577..651B,2008A&A...490..571F}. More recently, there has also been renewed interest in the volumetric Schmidt-type SF law which considers the volume-densities of the gas and SFR \citep{2012ApJ...745...69K,2014ApJ...782..114E,2019A&A...622A..64B}. Indeed, \citet{2019A&A...622A..64B} found `an unexpected and tight' correlation between the volume densities of SFR and atomic gas, highlighting the important role of the \HI\ gas, especially in low-density and low-metallicity environments. 

This study, a prelude to a second paper on star formation thresholds in nearby galaxies on sub-kpc scales (Paper 2 hereafter), is aimed at characterizing a sample of 228 resolved galaxies in terms of their scaling relations. 
The relations range from global properties of the gas disk such as the HI size-mass relation \citep{1997A&A...324..877B,2016MNRAS.460.2143W} to correlations between the gas disk, the assembled stellar disk and the star formation rate, e.g. the HI mass-stellar luminosity correlation and star forming main sequence. Such scaling relations provide us with insights into the connections between the instantaneous star formation and the gas reservoir, enable us to study the efficiency of gas consumption during star formation at various cosmic epochs, as well as to investigate mechanisms for gas replenishment such as gas accretion (e.g. \citealt{2005MNRAS.363....2K,2007AJ....134.1019O,2008A&ARv..15..189S}). %

This paper is organized as follows: Section \ref{thesample} presents the sample and data used, while in Section \ref{themethods} we discuss the methods used to estimate the galaxy properties of gas mass, stellar mass and star formation rates. In the same section, we also present new \HI\ intensity maps and velocity fields for 228 galaxies from the WHISP survey (Westerbork survey of \HI\ in Irregular and SPiral galaxies, \citealt{1996A&AS..116...15K,2001ASPC..240..451V}) as well as the methodology used to derive them. In Section \ref{therelations}, we present and discuss the derived global scaling relations. While \ref{hi_diams_sec} and \ref{mhivslight} focus on the global \HI\ properties and the distributions of gas mass and mass-to-light ratio as functions of the 3.4 $\mmu$ stellar luminosity and surface brightness, \ref{section:sfms} and \ref{section:sfrmhi} focus on a sub-sample of 180 galaxies that are detected in W3 (11.6 $\mmu$, which is sensitive to the heating of the interstellar medium by star formation) to study the galaxy SF main sequence relation and the Kennicutt-Schmidt relation. We summarize our findings in Section \ref{theconclusion}.

\section{Sample and Data} \label{thesample}
\begin{figure}
\centering
\includegraphics[scale=0.3]{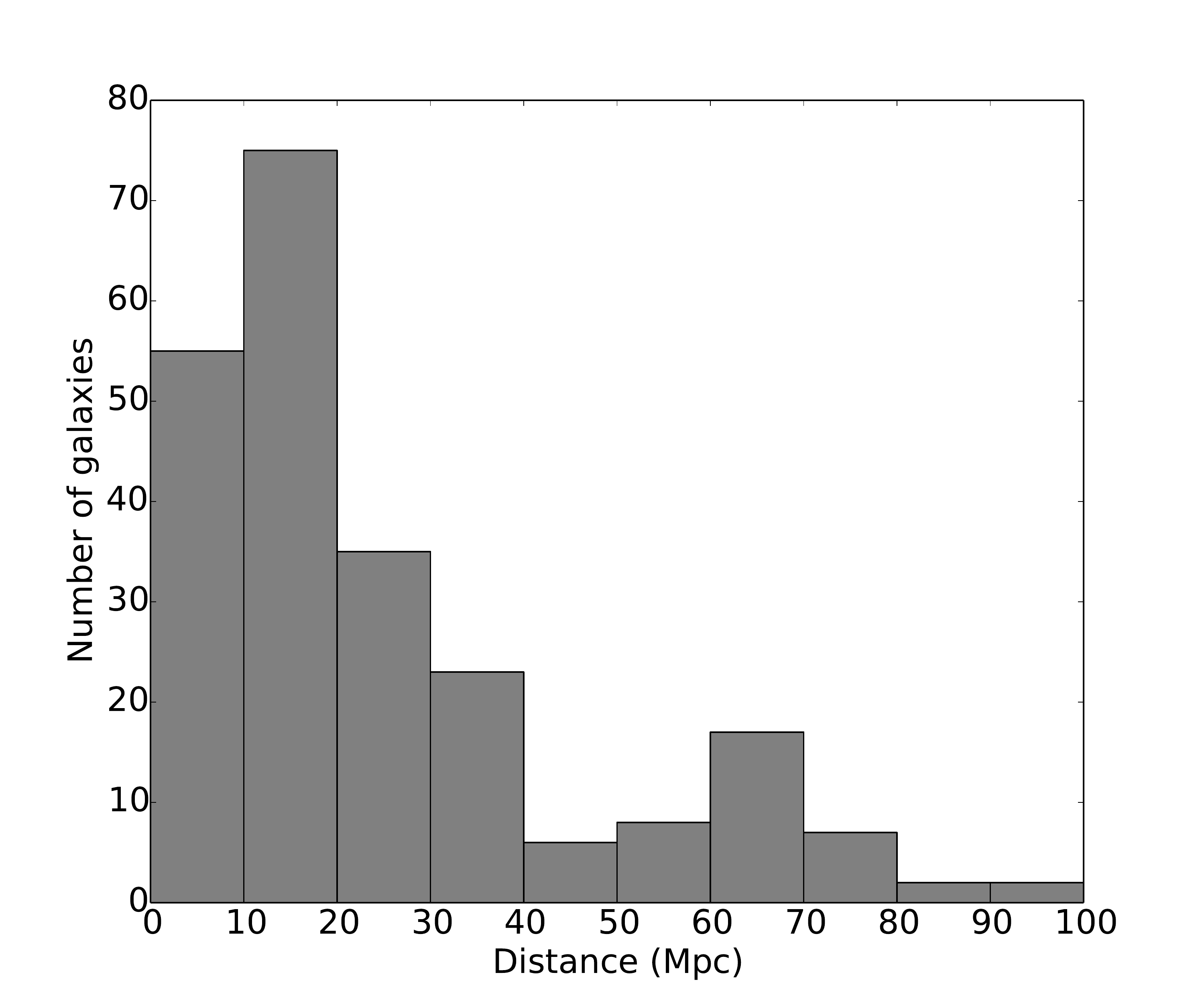}
\caption[Distance distribution of the WHISP data]{Distance distribution of the main sample used in this study. For these 228 galaxies, new data products have been generated (see Section \ref{themethods})} 
\label{dist_hist}
\end{figure}

The sample used in this study was chosen from the WHISP $30''$ resolution data which have higher sensitivity to low \HI\ column density structures than the full resolution (14" x 14" /$\mathrm{\sin\delta}$) data. Also, given the distance range 5~Mpc - 30~Mpc, where 70$\%$ of the sample falls (see Figure \ref{dist_hist}), the angular resolution of ~$30''$ enables us to investigate physical scales of $\sim$1~kpc. (see e.g. \citealt{2002AJ....124.2581S,2008AJ....136.2846B,2010ApJ...722L.127O,2011ApJ...735...63L,2013ApJ...779....8K,2019MNRAS.483..931E} for studies of the star formation relation at sub-kpc scales).

\begin{figure}
\centering
\includegraphics[scale=0.3]{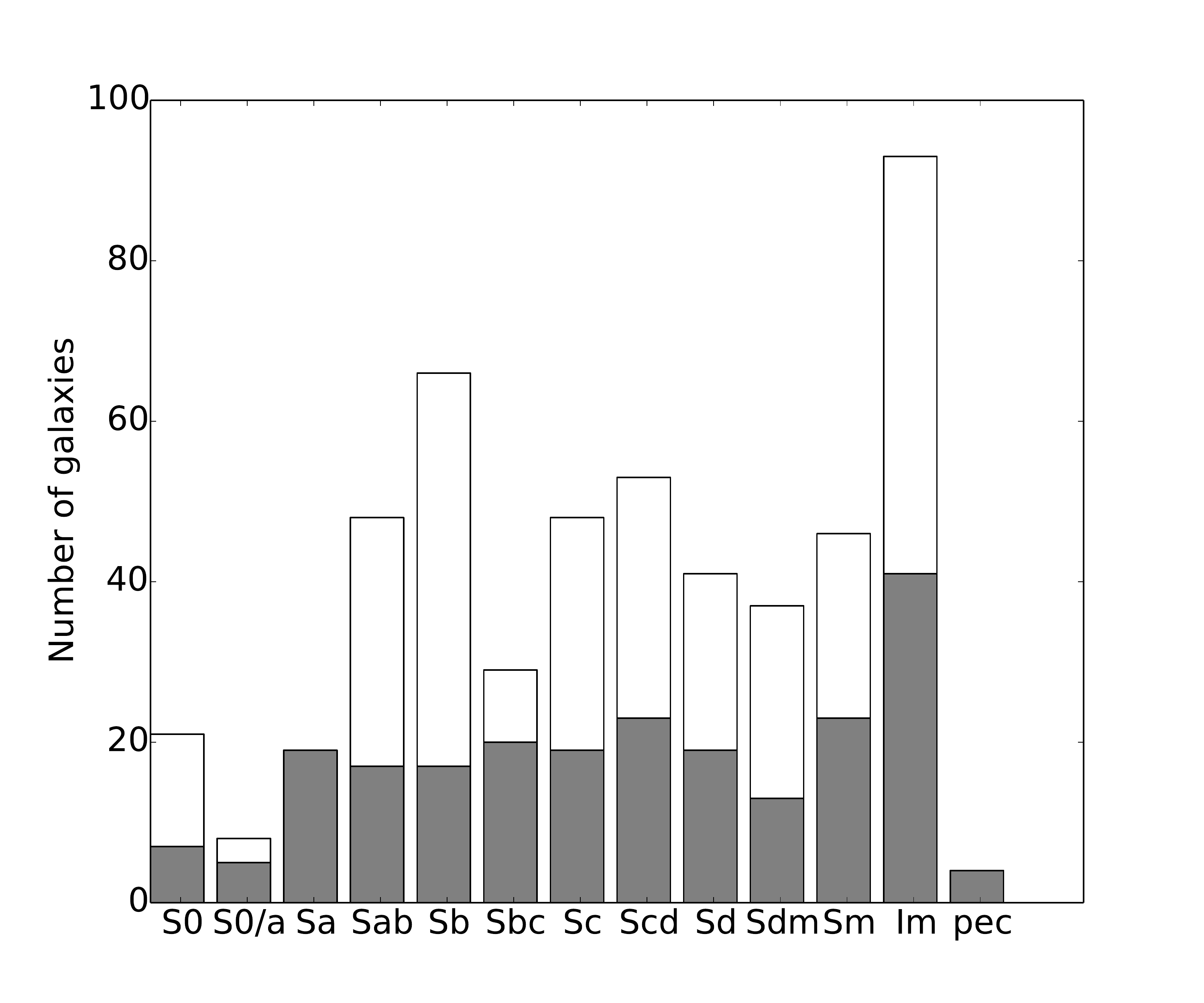}
\caption[Morphological distribution of the WHISP data]{Distribution by morphology of the WHISP sample. The main sample used in this study (shaded area, 228 galaxies) fairly represents the entire WHISP sample (background unshaded histogram). The morphologies were obtained from the Third Reference of Bright Galaxies (RC3.9, \citealt{1991rc3..book.....D}) via the NASA/IPAC Extragalactic Database (NED).}
\label{morph_hist}
\end{figure}

\begin{figure*}
\includegraphics[scale=0.2]{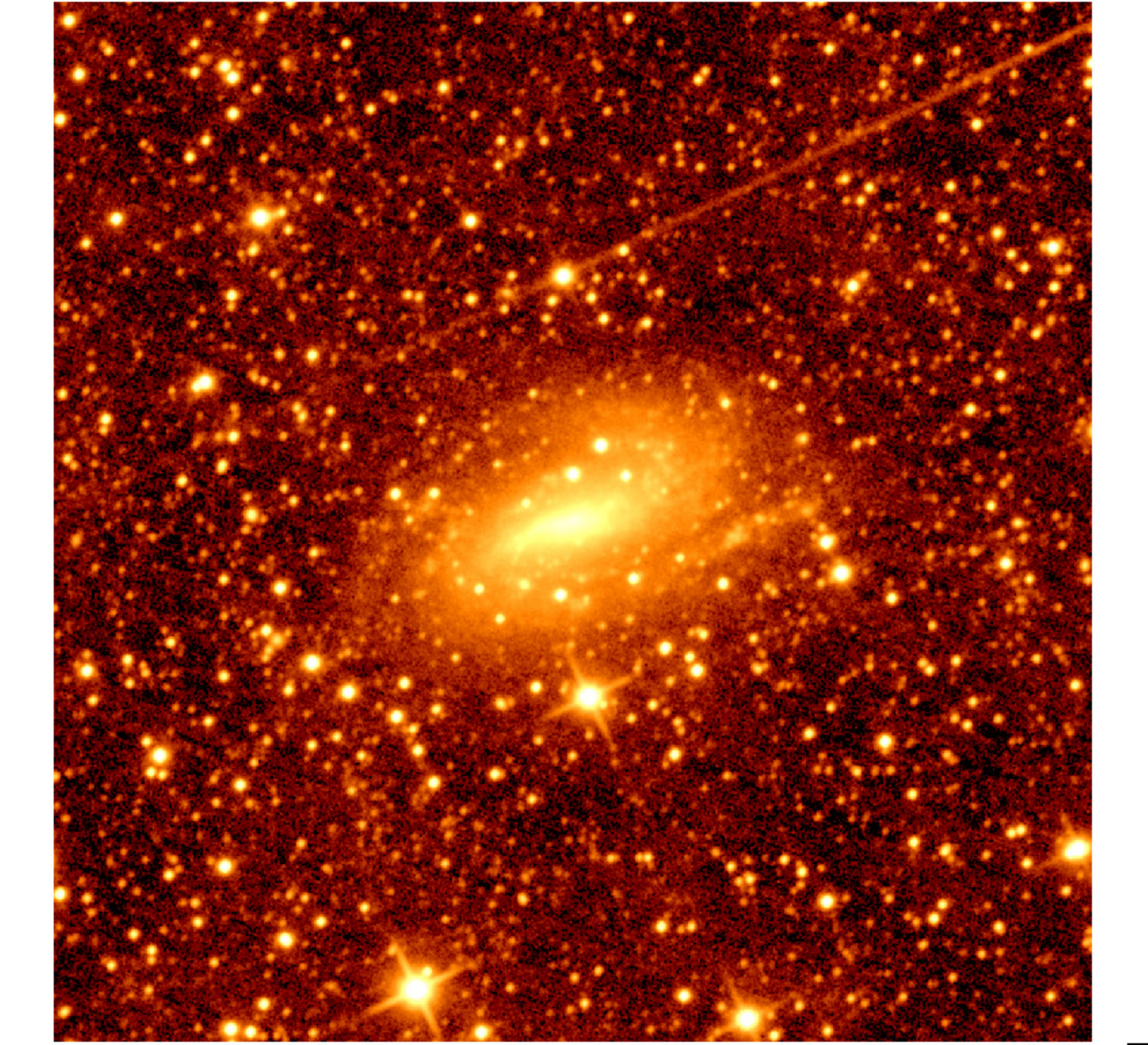}
\includegraphics[scale=0.2]{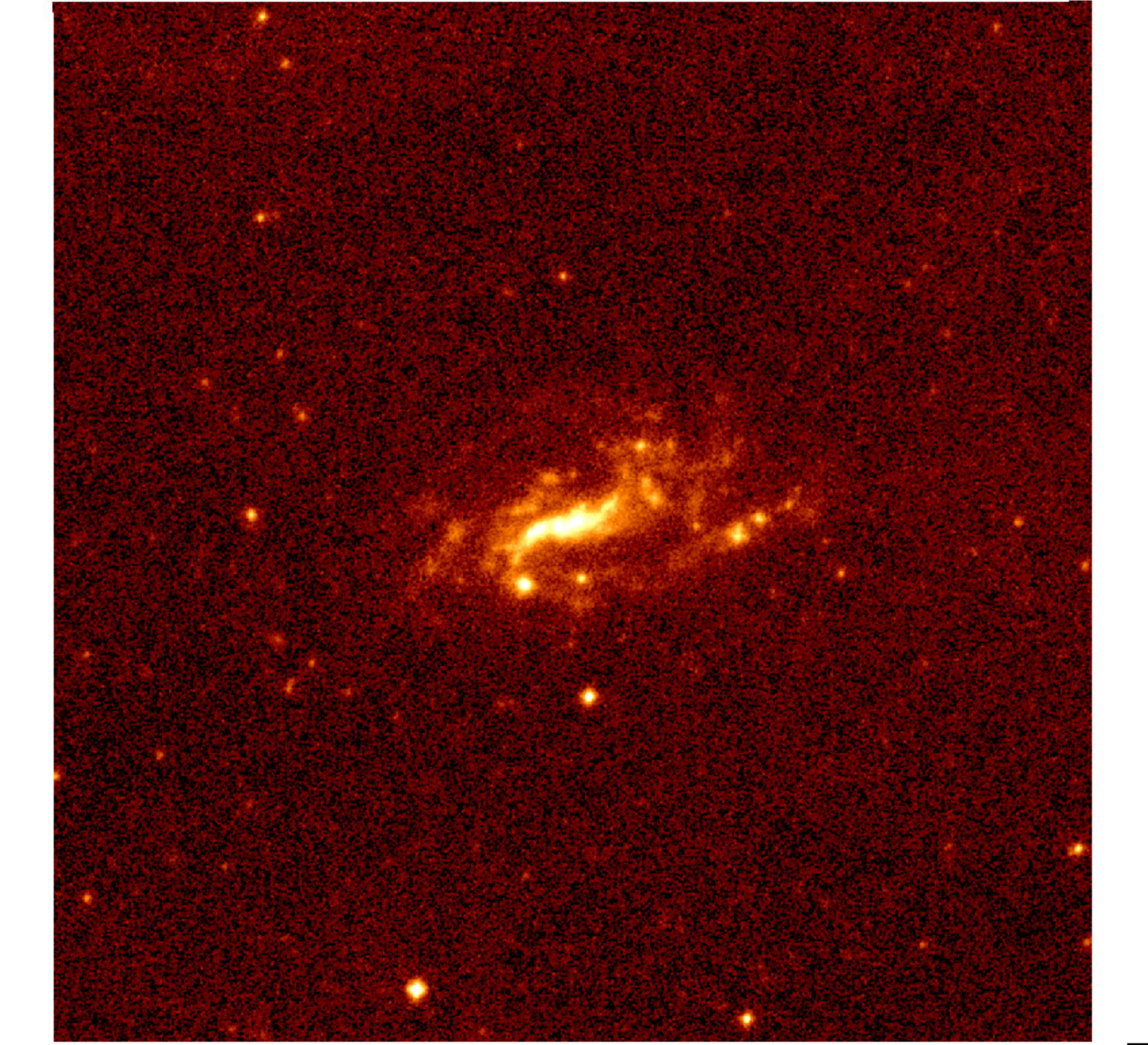}
\includegraphics[scale=0.25]{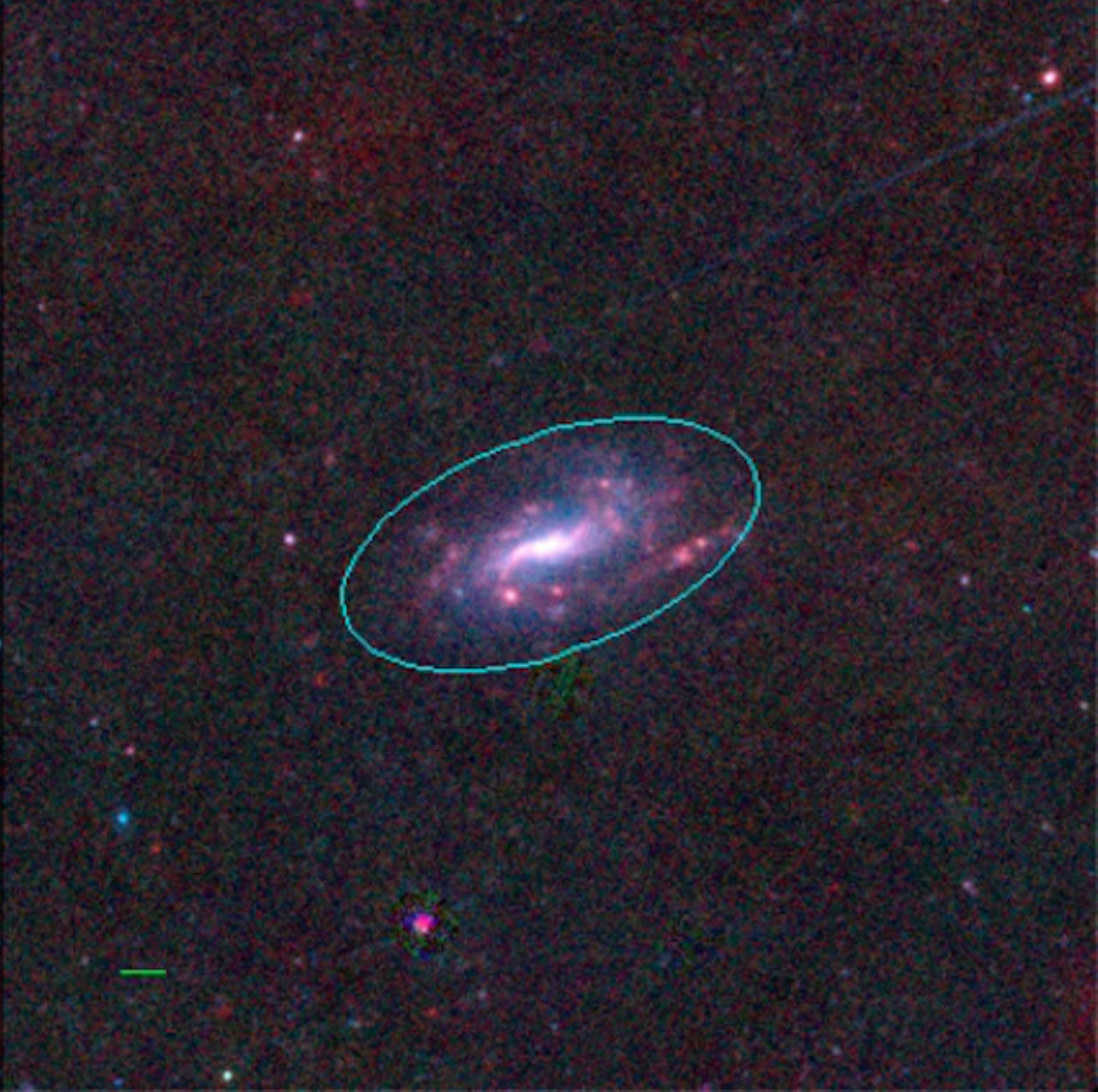}
\caption{WISE images of UGC1913 (NGC0925) in 3.4 $\mmu$ (\textit{left}), 12 $\mmu$ (\textit{middle}) and three-color view (3.4 $\mmu$, 4.6 $\mmu$ and 12 $\mmu$), with foreground stars removed, leaving only the target galaxy -\textit{right panel}. The green scale bar shows an angular scale of 1 arcmin (2.6 kpc).}
\label{u1913stamps}
\end{figure*}

Infrared data for the sample were obtained from the WISE Extended Source Catalogue \citep{2019arXiv191011793J} which constructed native resolution mosaics and carried out source characterization designed for large galaxies.  WISE mid-IR imaging is sensitive to stellar mass (in the short-wavelength bands) and ISM activity (long wavelength bands). Galaxies in close interactions (close enough for the interaction to be seen in the infrared images) or mergers as well as poor data and non-detections were excluded. These data were visually inspected to ensure there were no disturbances on the target’s IR flux by bright foreground stars. Wherever the masking or subtraction of these stars adversely compromised the disk of the galaxy, that target was dropped from the sample. Figure \ref{bad_phot} illustrates how this was achieved.
These criteria resulted in a sample of 228 galaxies. Since the resultant sample was determined by data quality, it is not biased by morphology and can be taken as a fair representation of the complete WHISP sample. The morphologies in this sample are plotted in Figure \ref{morph_hist} against the complete WHISP sample presented by \citet{2001ASPC..240..451V}. 
When investigating the global star formation properties (Sections \ref{section:sfms} and \ref{section:sfrmhi}), only the galaxies detected in the W3 ($\mathrm{12\ \mmu}$) band were chosen, leading to a sub-sample of 180 galaxies. Figure \ref{u1913stamps} shows images of the spiral galaxy UGC1913 in the WISE bands at $\mathrm{3.4\ \mmu}$ and $\mathrm{12\ \mmu}$. 

\section{Data Products}\label{themethods}
The WHISP survey provides both data cubes and moment maps; however, we derived new maps by employing a strict noise-rejection technique in order to foster more robust measurements of the global \HI\ properties. In this section, we describe the methods used to derive \HI\ intensity maps and velocity fields.
\subsection{\HI\ intensity maps}\label{HImethods}
$30''$ resolution data cubes were obtained from the \href{http://wow.astron.nl/}{public database} of the WHISP\footnote{\label{w}https://www.astro.rug.nl/$\sim$whisp/} survey. The cubes were corrected for primary beam attenuation using the \textsc{MIRIAD}\footnote{\label{+}http://www.atnf.csiro.au/computing/software/miriad/} \textit{linmos} task. We describe below the process\footnote{\label{p}Note that python scripts were written for each of these steps for full control of the process.} we implemented to generate a new set of high-quality HI maps.

\begin{figure}
\raggedright
\begin{minipage}{8cm}
\includegraphics[scale=0.35]{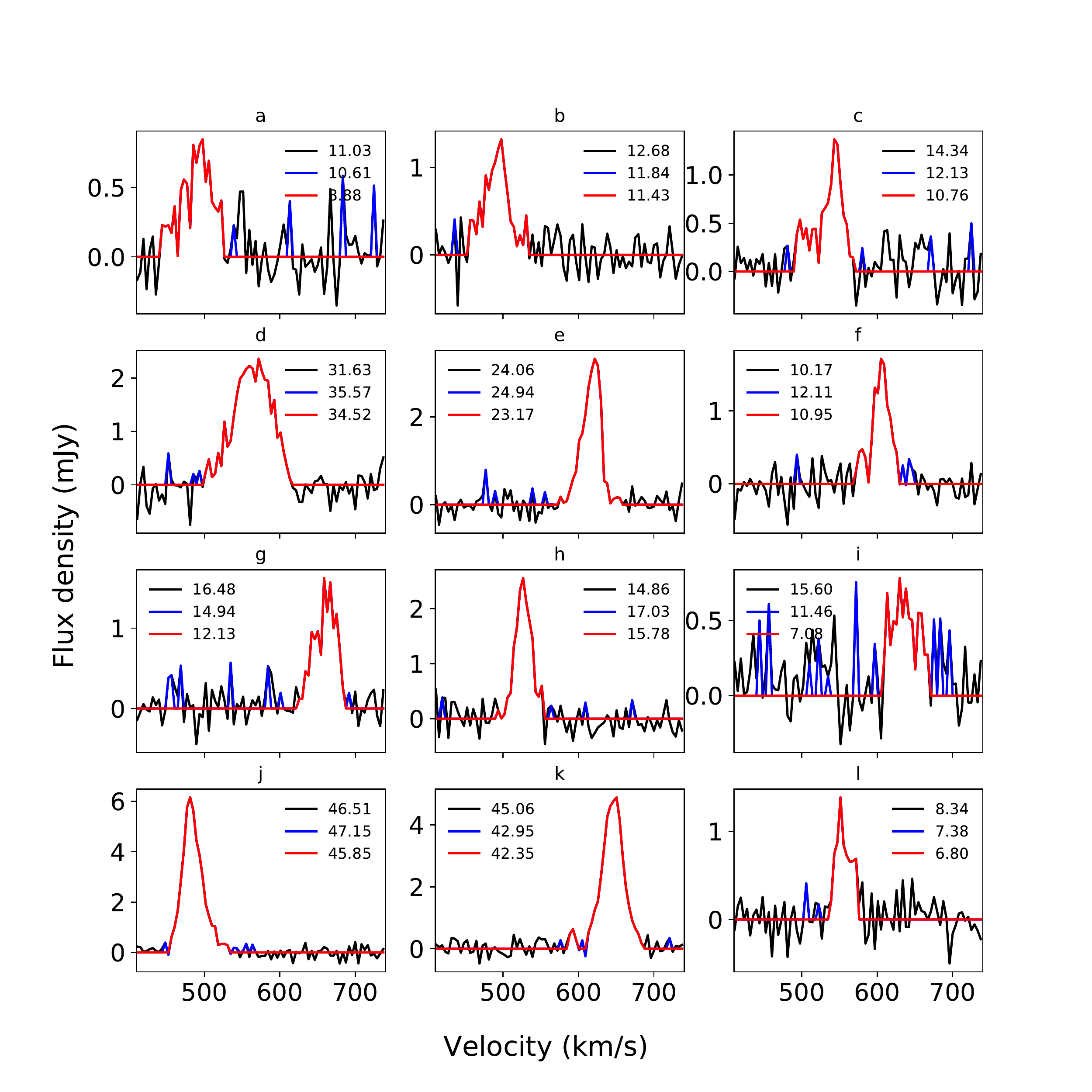}
\caption[Masking high noise peaks]{Line profiles at different pixel positions, drawn from the data cube of UGC1913, to illustrate the multi-channel-peak criterion. The black shows profiles from the original cube. The blue shows profiles from the sigma clipped cube which survived the 2$\sigma$ clipping. These were removed by using the multi-channel-peak criterion whose resulting profiles are shown in red. The total flux under each profile is indicated by the corresponding legend. The pixel positions of the line profiles are indicated in Figure \ref{mom0_example}.} 
\label{multi-peak_criterion}
\end{minipage}
\end{figure}

\begin{figure*}
\centering
\subfloat[]{\includegraphics[scale=0.4]{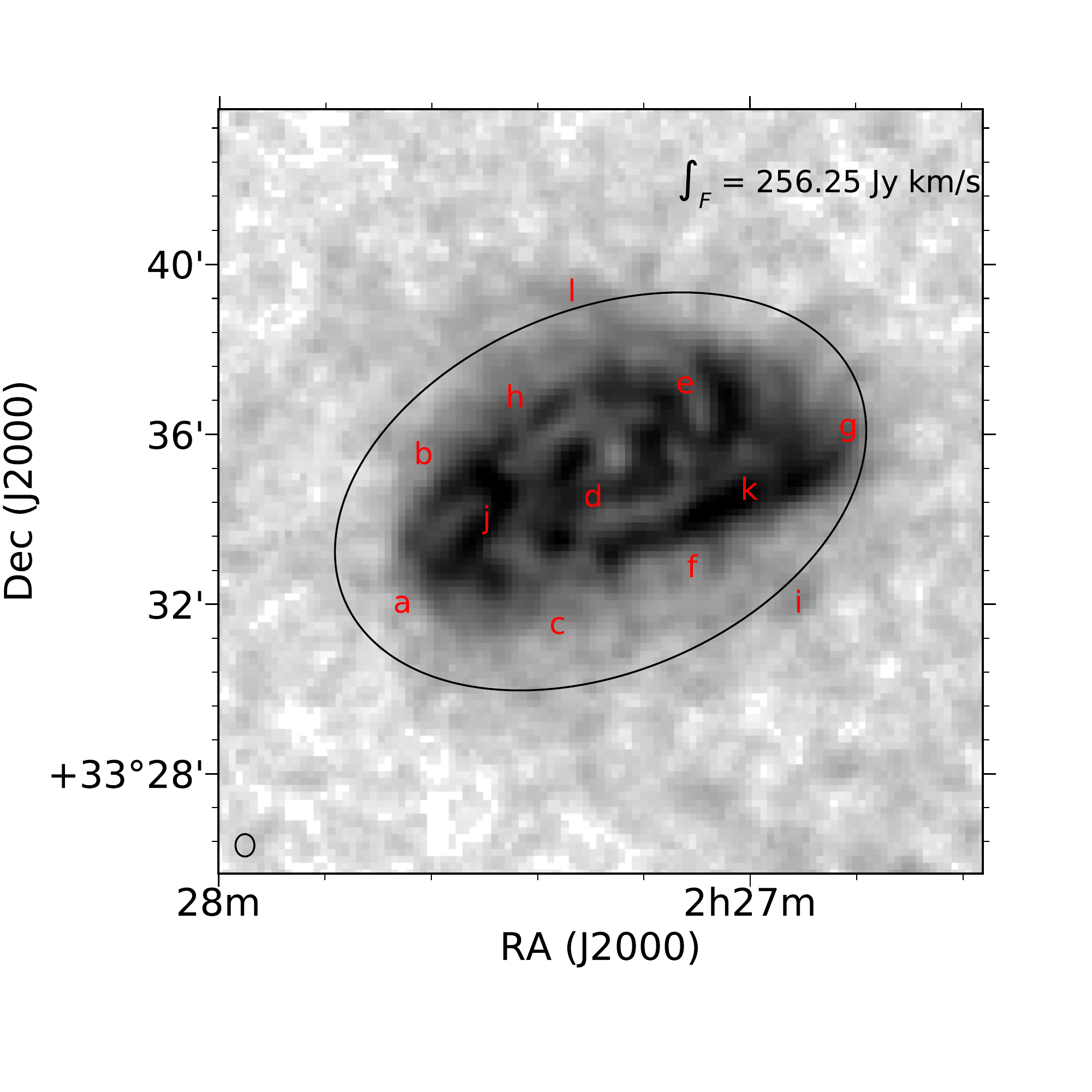}}
\subfloat[]{\includegraphics[scale=0.4]{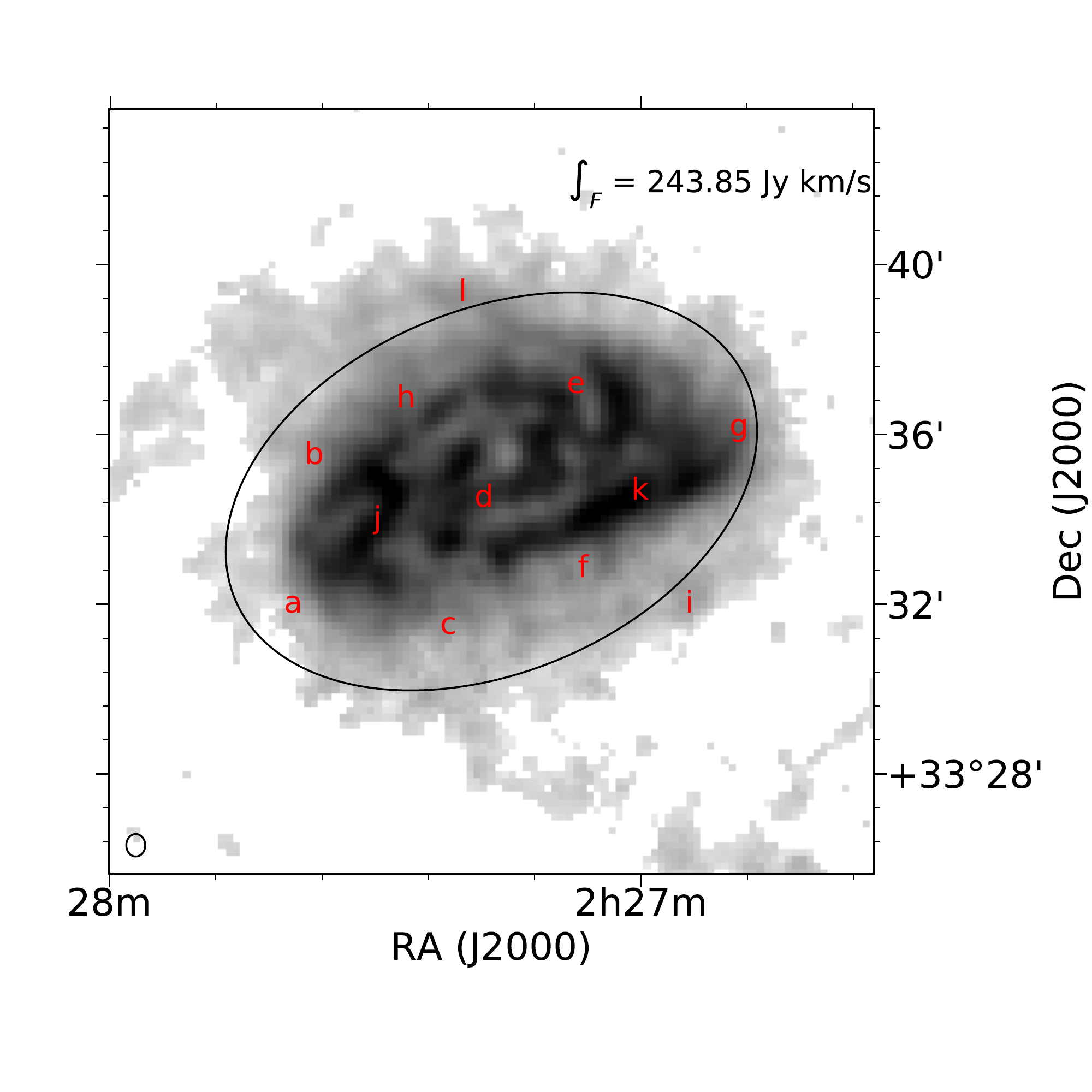}}\\
\caption[]{\HI\ intensity maps of the late-type spiral galaxy UGC1913. Panel (a) shows the result of adding up the emission in the sigma clipped cube, while (b) shows the result when the same cube is further processed using the multi-channel-peak criterion described in text. The extra step gets rid of sharp noise peaks which may be spatially coincident with the galaxy but separated in velocity space. The flux in the region enclosed by the black ellipse is shown in the upper right corner of each image. Letter labels indicate positions of the line profiles in Figure \ref{multi-peak_criterion}. The beam size ($\sim 30'' \times \sim 30''$) is shown in the lower left corner of each panel. These data products will be made available upon request.} 
\label{mom0_example}
\end{figure*}

The cubes were first smoothed to $60''$. For each smoothed cube, the RMS of the flux in a line-free channel was measured. Then, all flux in the cube below the 2$\sigma$ level was blanked while the flux above was marked as \textsl{True}. The resulting masks were in turn applied to the original $30''$ beam cubes to produce sigma-clipped cubes. 
To remove high noise peaks which survived the sigma clipping, a further step (multi-channel-peak criterion, see \citealt{2005A&A...442..137N}) was taken, where line profiles through each pixel position were inspected. Peaks detected in three or more channels were considered to be galaxy flux, but otherwise flagged as noise. Figure \ref{multi-peak_criterion} shows examples of line profiles (from the data cube of UGC1913) showing this process. From the resulting cube, the \HI\ intensity map was obtained by summing up the intensities from all the emission containing channels. The intensity maps from the noise clipping process before and after application of the multi-channel-peak criterion are shown in Figure \ref{mom0_example}. This same cube was used in deriving the global profiles.\\

In addition to the intensity distribution maps, we have derived new velocity field and dispersion maps for the WHISP sample by fitting third order Gauss-Hermite polynomials to the individual line profiles. The dispersion maps are used to model the disk gravitational instabilities in Paper II while the velocity fields will be used to derived rotation curves and study the Tully-Fisher relation in a later paper. These data products (\HI\ intensity maps and velocity fields and dispersion maps) will be made available upon request.
\subsection{\HI\ mass}
\begin{figure}
\centering
\subfloat[]{\includegraphics[scale=0.35]{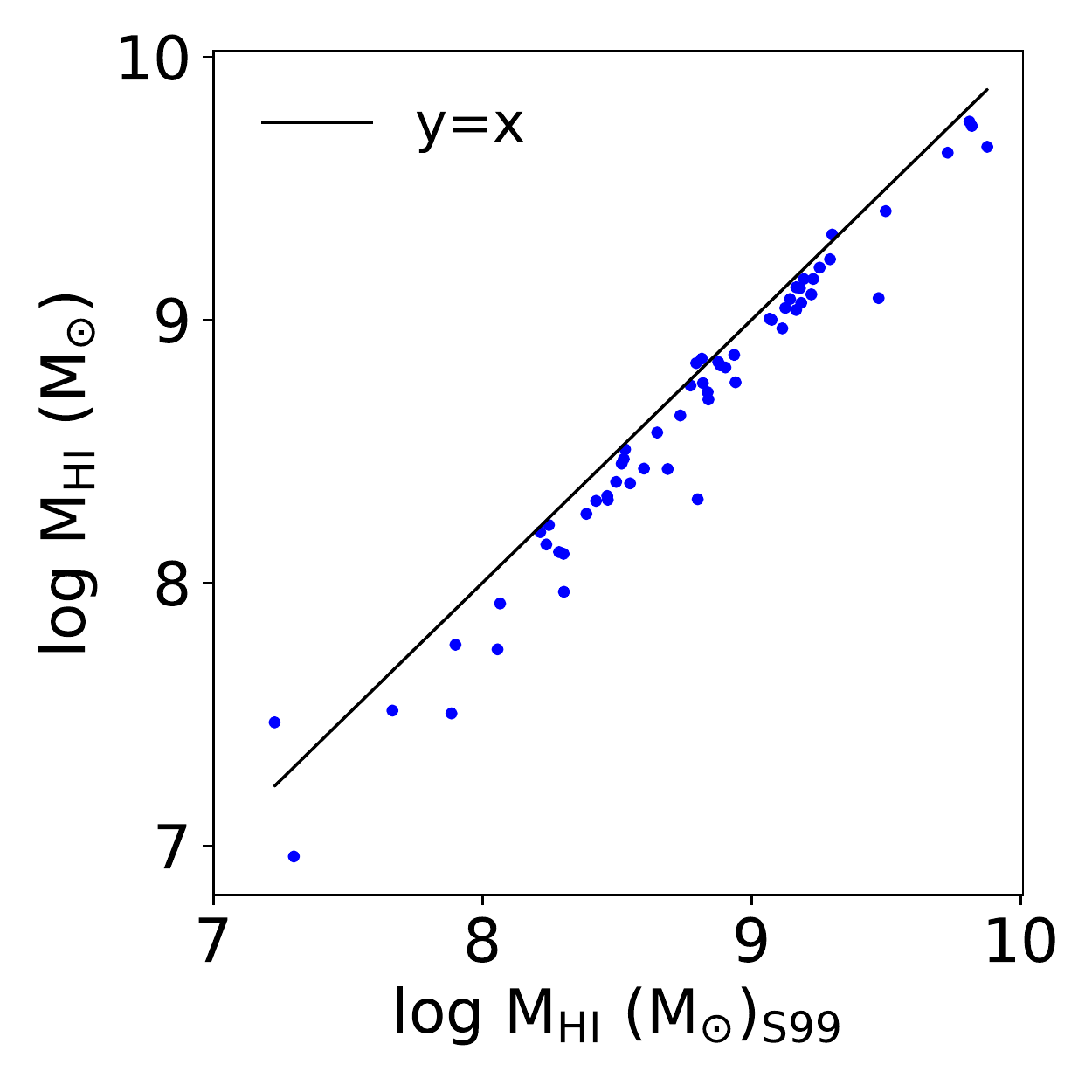}}
\subfloat[]{\includegraphics[scale=0.35]{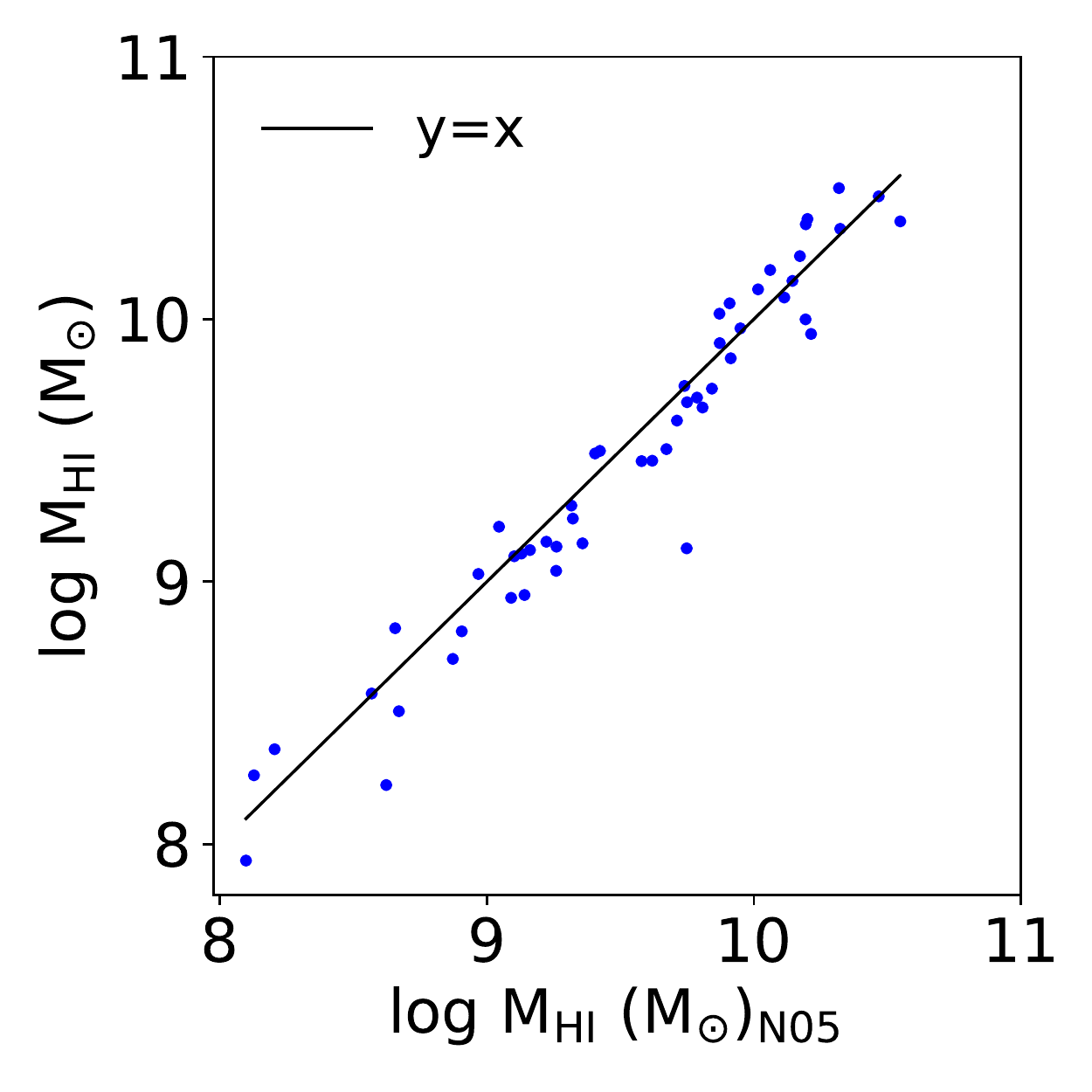}}
\caption[Comparison of masses]{Comparison of mass estimates from the literature (x-axes) with this study (y-axis). S99 (a) retrieved higher fluxes from the WHISP cubes because they did not carry out the multi-peak criterion for noise removal, which N05 (b) carried out and had fluxes in agreement with ours. Note that N05 had a sample of 68 early-type spirals while S99 had sample of 73 late-type dwarfs. Our sample of 229 had 52 and 60 galaxies in common with N05 and S99. All the mass estimates shown on these plots are total integrated \HI\ masses.}
\label{mass_comparison}
\end{figure}

The total \HI\ mass was calculated from the total \HI\ flux as,
\begin{equation}
\mathrm{M_{\HI}(M_\odot)\ =\ 2.36 \times 10^5 \times D^2 (Mpc)\times \int{S_\nu}\ dv\ (Jy\ km\ s^{-1})},
\label{MHIequation}
\end{equation}
\citep{Brinks2006} where D is the distance, and $\mathrm{\int{S_\nu}\ dv}$ is the total \HI\ flux integrated over the global profile. 
Figure \ref{mass_comparison} shows comparisons of the masses derived in this study with masses derived by two other studies that used WHISP data. These are \citet{1999PhDT........27S}(hereafter S99) and \citet{2005A&A...442..137N}(hereafter N05), with whom we had 60 and 52 galaxies in common, respectively. The masses were compared after synchronizing the distances across the samples (see Table \ref{tab_props_with_dists} for adopted distances). The masses in this study are systematically lower than those of S99 by 0.12 dex on average. This is expected because S99 did not carry out the extra noise handling by the multi-channel-peak criterion hence having higher fluxes. On the other hand, N05 employed the multi-channel-peak criterion when generating their data products which explains why their mass estimates are in general agreement with ours. The derived total flux and mass as well as linewidths and systemic velocities for all galaxies in the sample are listed in Table \ref{tab_props_with_dists}.

\subsection{\HI\ surface density}
Maps of \HI\ surface density were obtained from the integrated flux maps as below, following the methods of \citet{Brinks2006,2008AJ....136.2782L}. The \HI\ maps were converted into column densities using the equation;
\begin{equation}
\mathrm{N_\HI\ (cm^{-2})\ =\ 1.823\times10^{18}\times\frac{\sum_{\nu} S_{\nu}}{1.65\times 10^{-3}\times\Delta\alpha \Delta\delta}\times \Delta V},
\end{equation}
where $\mathrm{N_\HI}$ is the column density map in atoms $\mathrm{cm^{-2}}$, $\mathrm{\Delta\alpha \ and\ \Delta\delta}$ are the major and minor axes of the synthesized beam, $\mathrm{\Delta V}$ is the velocity resolution in $\kms$, and $\mathrm{\sum_{\nu} S_{\nu}}$ is the integrated \HI\ map in  $\mathrm{mJy/beam}$. The factor in the denominator converts the intensity map into a brightness temperature map.\\
The column density maps were converted into surface density maps using the equation;
\begin{equation}
\mathrm{\Sigma_{\HI}(M_\odot\ pc^{-2})\ =\ \frac{N_\HI}{1.248\times10^{20}}\ \cos i},
\end{equation}
which accounts for inclination.
Radial profiles were obtained from the inclination corrected surface density maps using the \textsc{MIRIAD} task \textit{ellint} and used to derive the \HI\ diameters at $\mathrm{\Sigma_{\HI}\ =\ 1\ M_\odot\ pc^{-2}}$. The threshold value of $\mathrm{1\ M_\odot\ pc^{-2}}$ was chosen in keeping with the literature (see, for example, \citealt{1997A&A...324..877B,2001A&A...370..765V,2005A&A...442..137N,2016MNRAS.460.2143W})

\newpage
\subsection{W1 stellar mass estimation}\label{stellarmasssection}
The  WISE $3.4\mmu$  (W1) band is a reliable measure for the stellar mass of galaxies because it traces the evolved stellar population which comprises the bulk of the baryonic mass in galaxies \citep{2013AJ....145....6J,2014ApJ...788..144M}. Furthermore, the near infrared does not suffer as much extinction as optical bands. We calculated stellar masses for stellar disks from the W1 absolute magnitude. The stellar disks were defined as the region enclosed by the 1$\sigma$ isophote \footnote{\label{sp}$\sim$23 \marc in Vega units} in the WISE W1 images. The stellar mass empirical relations of \citet{2014ApJ...782...90C} were used;
\begin{equation}
\mathrm{L(L_\odot)}\ =\ 10^{-0.4(M\ -\ M_{sun})},
\end{equation}
\begin{equation}
\mathrm{M_\star(M_\odot)}\ =\ \mathrm{\Upsilon_*^{[3.4]}\ \times \ L},
\end{equation}
\begin{equation}
\mathrm{\log\ \Upsilon_*^{[3.4]}\ =\ -0.17\ -\ 2.54(W1-W2)}
\end{equation}

where L is the W1 in-band luminosity, $M$ is the absolute magnitude in the W1 band, $M_{sun}=3.24$ is the absolute magnitude of the sun in the W1 band \citep{2008AJ....136.2761O,2013AJ....145....6J} and $\mathrm{\Upsilon_*^{[3.4]}}$ is the mass to light ratio. 
Limits were placed on the W1-W2 color ($\mathrm{-0.05 \leq W1-W2 \leq 0.2mag}$) to minimize contamination by AGN light in W2 and  unphysical blue colors (due to the lower sensitivity of the W2 band). Below the lower limit, the M/L was set to 0.21, and to 0.91 above the upper limit \citep{2014ApJ...782...90C,2019arXiv191011793J}. Any dwarf galaxies which were not detected in W2 or had W1-W2 colors with S/N $\leq$ 3 were assigned a constant M/L of 0.6 \citep{2018MNRAS.473..776K}.
%
The derived stellar masses for our sample are presented in Table \ref{tab_ks_samp}. Note that the W1 band is sensitive to light from warm dust, PAH molecules (3.3 $\mmu$) and AGB stars, all of which can exaggerate the W1 flux \citep{2012ApJ...744...17M,2015ApJS..219....5Q,2018MNRAS.474.4366P}. The independent component analysis correction for non-stellar contamination \citep{2012ApJ...744...17M} was not carried out, and hence the stellar masses quoted here should be taken as upper boundaries.

\subsection{W3 SFR estimation}
Measurements of Star formation were obtained from WISE W3 (11.6 $\mmu$) imaging. The W3 band is wide and spans PAH emission features at 7.7 $\mmu$, 8.5 $\mmu$ and 11.3 $\mmu$, Neon nebular emission lines [Ne II] at 12.8 $\mmu$ and [Ne III] at 15.6 $\mmu$, silicate absorption at 10 $\mmu$, as well as warm dust continuum \citep{2017ApJ...850...68C}. 
The 11.6 $\mmu$ band was chosen to derived SFR over the W4 (22.8 $\mmu$) band because of its superior sensitivity \citep{2013AJ....145....6J}. Furthermore, \citet{2019MNRAS.483..931E} have showed for M33 that the W3 band is a good monochromatic tracer of the total SFR (as typically traced by FUV + 24 um). \\
SFR's were derived from the spectral luminosity, $\mathrm{\nu L_\nu}$, following the empirical relations of \citet{2017ApJ...850...68C} who calibrated the WISE mid-IR to the total infrared luminosity;
\begin{equation}
\mathrm{\log(SFR)}\ =\ \mathrm{(-7.76\pm 0.15)\ +\ (0.889\pm 0.018)\ \mathrm{\log(\nu L_\nu)}},
\label{sfr_cal}
\end{equation}
with
\begin{equation}
\mathrm{\nu L_\nu(L_\odot)} \ =\ \frac{\mathrm{\nu_{w3}\ \times \ 4\pi D^2\ \times \ f_\nu}}{3.846\ \times \ 10^{26}},
\label{nulnu_map}
\end{equation}
where $\mathrm{\nu_{w3}}$ is the mid-band frequency of the W3 band, $\mathrm{f_\nu}$ is the W3 flux, $\nu L_{\nu}$ is the spectral luminosity in $\mathrm{L_\odot}$ and D is the distance in meters. With these prescriptions and WISE measurements, the uncertainty in the SFRs will be about 30\% \citep{2017ApJ...850...68C,2019arXiv191011793J}. \\
The adopted distances are listed in Table \ref{tab_props_with_dists} while the WISE photometry and derived SFR's are presented in Tables \ref{tab_wise_phot} and \ref{tab_ks_samp}.
\section{Scaling Relations}\label{therelations}
\subsection{\textit{The \HI \ mass-size relation}}\label{hi_diams_sec}
The \HI\ masses of galaxies and their diameters defined at an \HI\ surface density (\sighi ) of 1 \mpc\ are known to be tightly correlated. This relation has been observed to hold across three and five orders of magnitude in diameter and mass respectively \citep{2016MNRAS.460.2143W}, for morphologies ranging from early-type spirals to dwarfs and irregulars \citep{1997A&A...324..877B,2001A&A...370..765V,2002A&A...390..829S,2005A&A...442..137N,2011PhDT.......327M,2014MNRAS.441.2159W,2016MNRAS.463.4052P}. The relations found by the different authors have slopes of 1.84$\pm$0.12 in log space. \citet{2002A&A...390..829S} and \citet{2005A&A...442..137N} derived this relation for 73 dwarf irregulars and 68 early-type spirals respectively in the WHISP sample. Here we present this relation for 228 WHISP galaxies, spanning from early-type spirals to dwarf irregulars and find the relation to be in agreement with previous studies within the errors. We find the tight relation below:

\begin{equation}
\mathrm{\log M_{\HI}\ =\ (1.95\pm0.03)\log D_{\HI}\ +\ (6.5\pm0.04)},
\label{eqn:dhimhi}
\end{equation}
which has a dex error of 3\% on the slope, 4\% on the intercept and a vertical scatter of 0.14 about the best-fit line.
The relation for our sample is in particularly good agreement with the relations of \citet{1997A&A...324..877B} and \citet{2016MNRAS.460.2143W} who had slopes of 1.96$\pm$0.04 and 1.98$\pm$0.01 respectively, and whose samples were complete comprising of all morphological types from early-type spirals to dwarfs as did our sample. Our slope is steeper than other studies by \citet{2001A&A...370..765V,2002A&A...390..829S,2005A&A...442..137N,2016MNRAS.463.4052P} mostly because their samples were less complete, consisting only of either spirals or dwarfs. Note that the \citet{2016MNRAS.460.2143W} study included 100 WHISP galaxies in their sample of 549.

\begin{figure}
\includegraphics[scale=0.5]{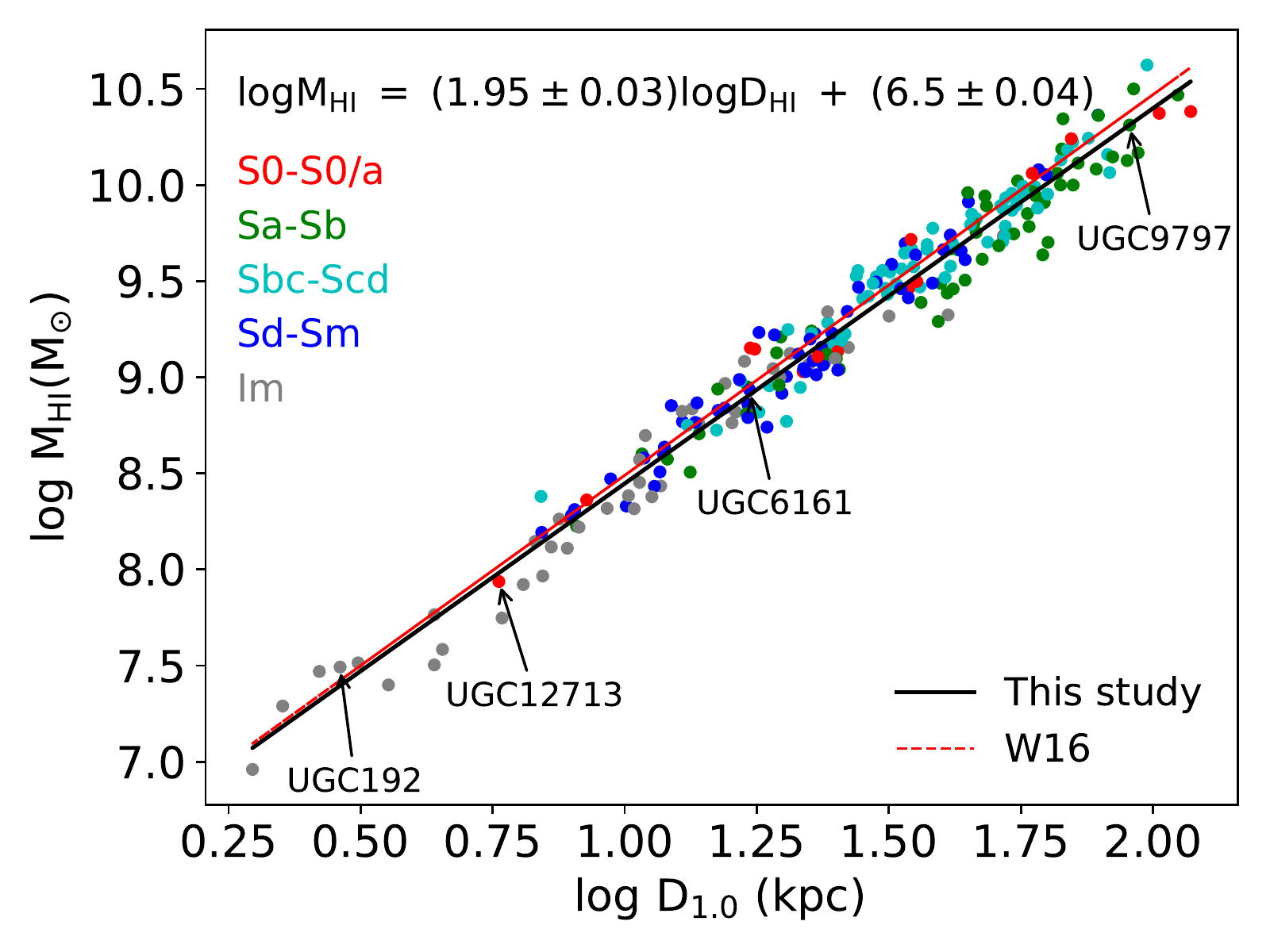}
\caption[\HI\ mass-size relation]{\HI\ mass versus \HI\ diameter for all galaxies in our sample. The data points are color coded by the morphology of the galaxies such that the red represents lenticulars (S0-S0/a), green represents early type spirals (Sa-Sb), cyan represents Sbc - Scd, blue represents late-type spirals (Sd-Sm) and the gray are the late-type dwarfs. A fit to the data is shown by the black solid line. The relation of \citet{2016MNRAS.460.2143W} (W16) is shown as a red dashed line. The \citet{1997A&A...324..877B}, \citet{2001A&A...370..765V} relations are very similar to the W16 relation. The \HI\ diameters were defined at the 1 \mpc\ level of the radial profiles while \HI\ masses were determined from the total flux in the global profiles. The mass-size relation of \HI\ disks is an indicator of a constant average \HI\ surface density. Note: One galaxy, UGC3826, had \sighi\ $\leq$ 1 \mpc\ across it's entire disk, and hence was not used when deriving the mass-size relation.} 
\label{dhi_mhi}
\end{figure}

Such a power law relates the approximate square of the diameter against the mass, giving a typical mean \HI\ surface density across galaxy disks \citep{1997A&A...324..877B}. That this relation is found consistent for samples across all morphologies and in different environments implies that \textit{the \HI\ disks regulate themselves to a typical mean value of surface density} \citep{2001A&A...370..765V,2005A&A...442..137N}. 
Having used analytical models and cosmological simulations for galaxies at z=0,  \citet{2019MNRAS.490...96S} attribute this property of galaxies to the general distribution of \HI\ in disks and its tendency to saturate due to the \HI\ - \molH\ phase transition, showing that even quenching and disk truncation will only change the relative position of a galaxy on the relation but will not completely remove it.
Figure \ref{dhi_mhi} is a plot of our data, overlaid with the relations of \citet{1997A&A...324..877B}, \citet{2001A&A...370..765V} and \citet{2016MNRAS.460.2143W}. Late-type dwarfs such as UGC192 (IC10) populate the low mass end of the relation while early-type spirals such as UGC9797 (NGC5905) are found at the high mass end. Exceptions exist such as an early-type at the low mass end which is the case for UGC12713, a dwarf spheroidal. However, this is  in keeping with the correlation that the smaller \HI\ disks contain lower \HI\ masses regardless of morphological type. 

\subsection{\textit{\HI\ mass vs stellar luminosity}}\label{mhivslight}
The relationships between the \HI\ mass and stellar disk properties of luminosity and surface brightness represent the scaling relations between the ISM which is a reservoir of fuel for star formation and the galaxy's underlying baryonic mass in the form of the old stellar population. The \HI\ mass derived here is the mass enclosed within the stellar disk, defined by a 1$\sigma$ isophote in the W1 image at surface brightness $\sim$ 23 \marc\ (Vega units), which incorporates more than 95\% of the total flux \citep{2019arXiv191011793J}. This gives us a direct comparison between the atomic gas reservoir for star formation spatially co-located with the stellar disk. 

\begin{figure*}
\centering
\includegraphics[scale=0.55]{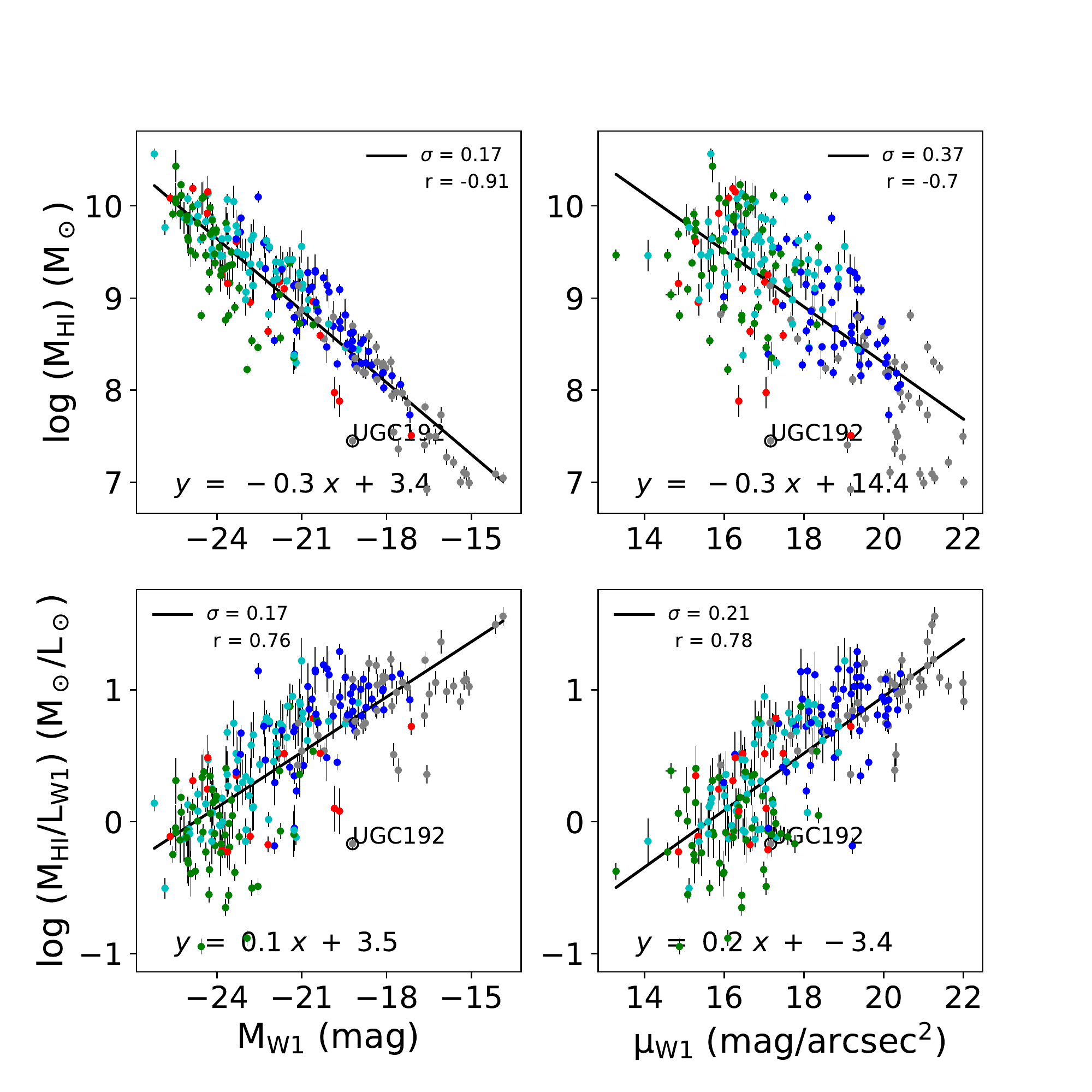}
\caption[Enclosed \HI\ mass vs Stellar luminosity]{Plots of \HI\ mass (upper panels) and \HI\ mass-to-light ratio (lower panels) versus 3.4 $\mmu$ absolute magnitude which is a direct measure of the stellar mass (left panels) and mean surface brightness over the stellar disk (right panels). Note that these are `enclosed quantities' measured within the stellar disks. The correlation coefficient (r) for each plot and the vertical scatter ($\sigma$) about the fitted line are shown in the upper left/right corners of the plots. For most points, the errors on the x axes are smaller than the data points. The plots are also color-coded according to morphological type such that red represents lenticulars (S0-S0/a), green represents early type spirals (Sa-Sb), cyan represents Sbc - Scd, blue represents late-type spirals (Sd-Sm) and the gray are the late-type dwarfs. }
\label{ml_mag}
\end{figure*}
Figure \ref{ml_mag} illustrates the relationships between the \HI\ mass contained within the stellar disk, and stellar disk magnitude and surface brightness. The plots are also color-coded according to morphological type such that red represents lenticulars (S0-S0/a), green represents early type spirals (Sa-Sb), cyan represents Sbc - Scd, blue represents late-type spirals (Sd-Sm) and the gray are the late-type dwarfs. 
 The `r' labels in the plots are the correlation coefficients while `$\sigma$' denotes the vertical scatter about the least squares fits to the data (solid line). All the plots show strong correlation coefficients between the properties. This is in agreement with the complete (although smaller) sample of \citet{2001A&A...370..765V}, as well as the sample of \citet{2016MNRAS.463.4052P}, although the \citet{2016MNRAS.463.4052P} study found a weaker correlation (0.64) between the $\mathrm{M_{\HI}}$ and stellar luminosity. 

From a morphological point of view, the late-type dwarfs have lower luminosities (intrinsically fainter) and higher surface brightness than the earlier type spirals. The intrinsic brightness decreases across the Hubble sequence from early-type spirals to late-type dwarfs.
We also see that the enclosed \HI\ mass decreases from early-types towards late-types while on the other hand the gas-star fraction increases towards the late-types. %

Note that our relations derived for the atomic gas enclosed within the stellar disk follow the same trend as those from previous studies that used integrated properties over the entire \HI\ disk. However, there is clearly less scatter in our relation between $\mathrm{M_{\HI}}$ and $\mathrm{M_{W1}}$, as would be expected, because we are comparing spatially co-located properties in the stellar disk.  

\begin{figure*}
\includegraphics[scale=0.14]{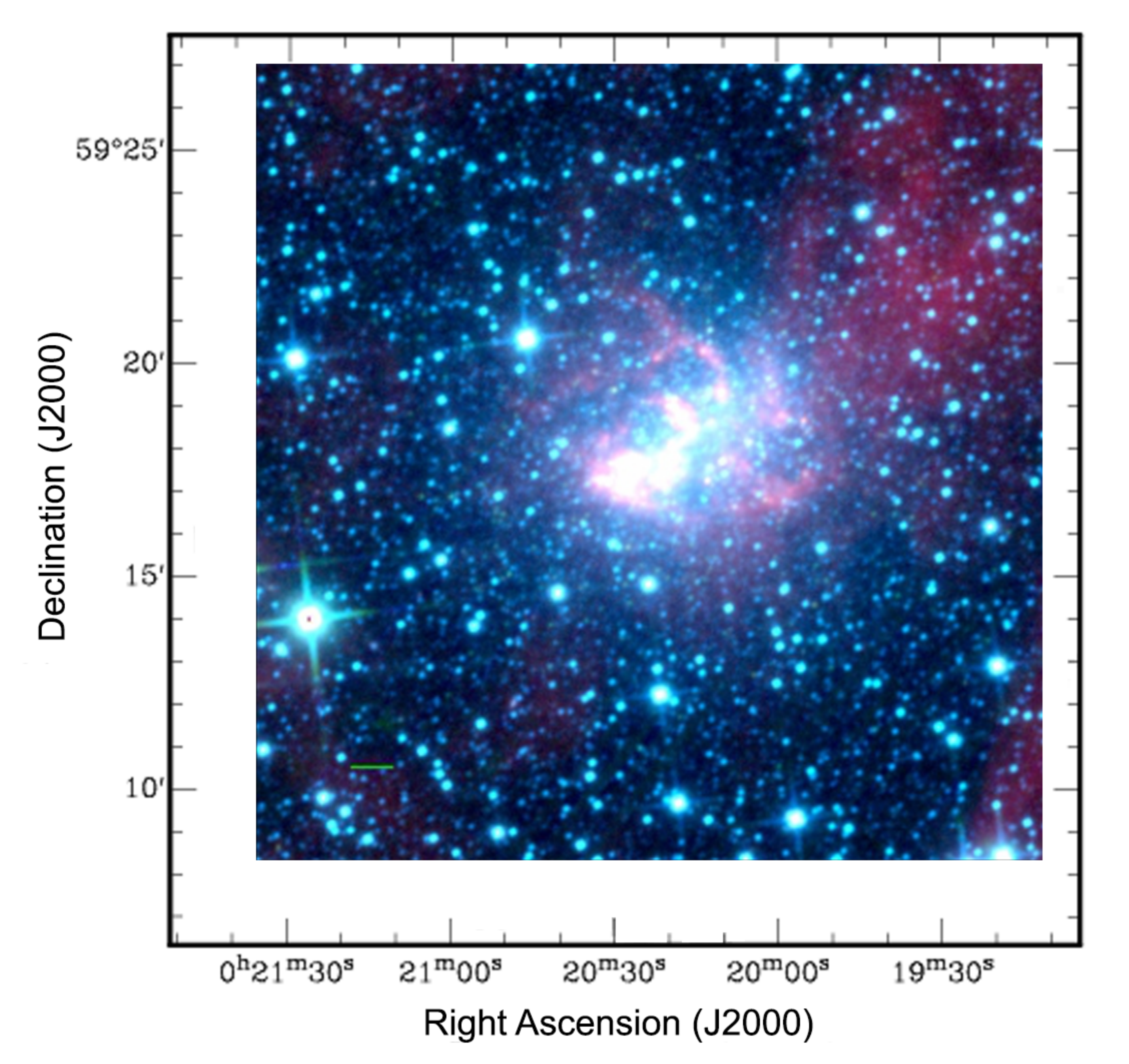}
\includegraphics[scale=0.14]{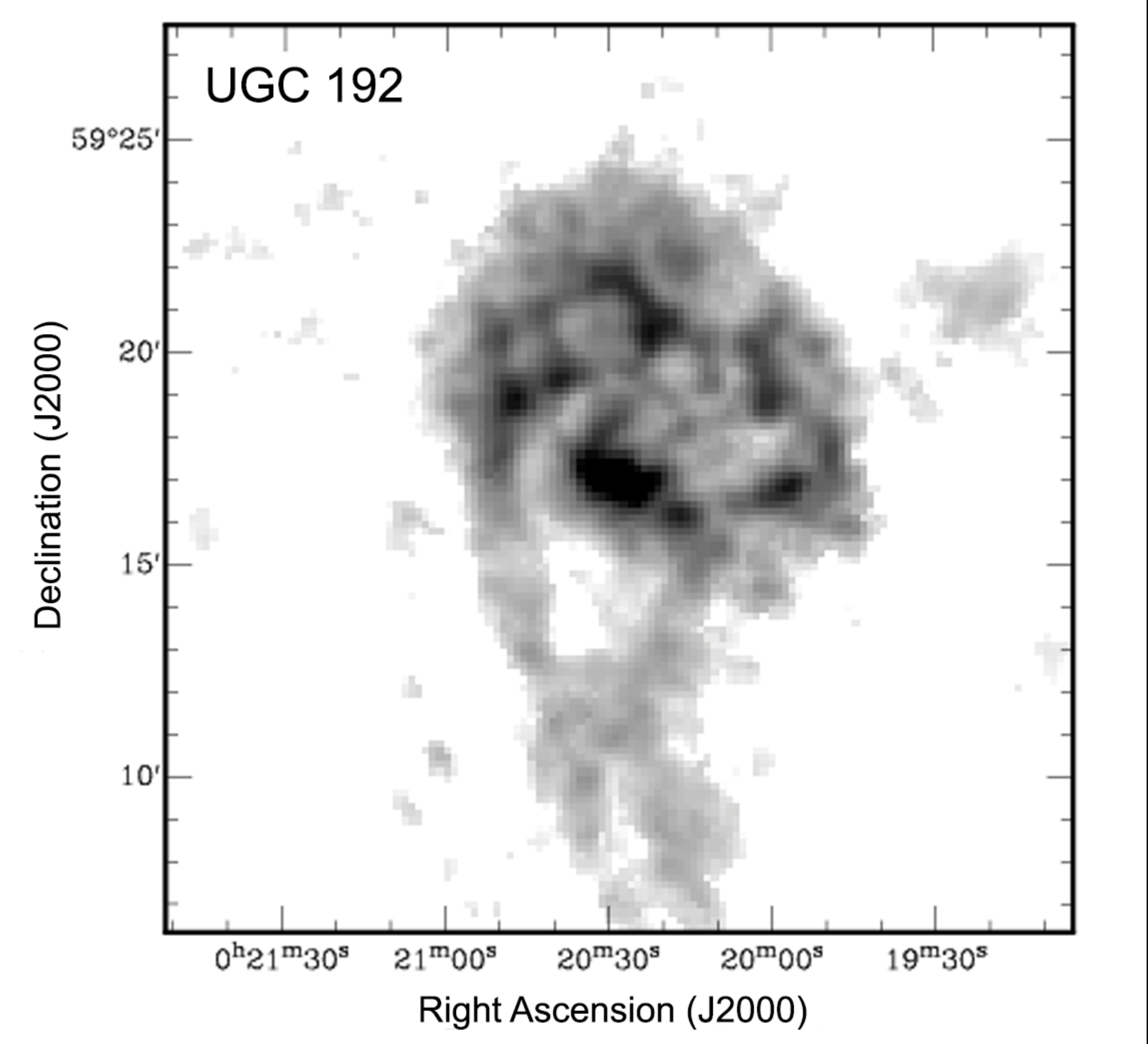} 
\includegraphics[scale=0.14]{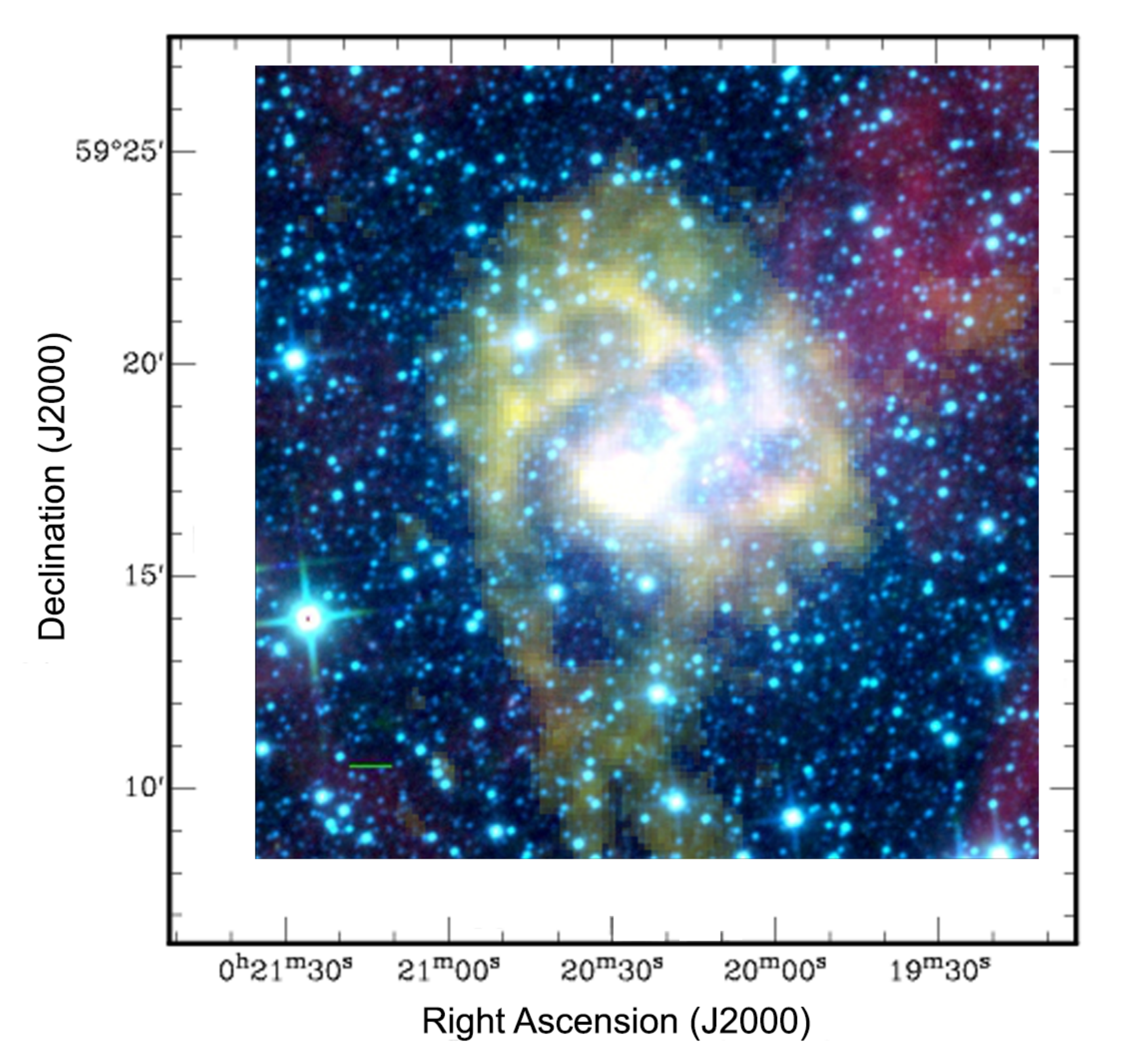}
\caption[UGC192 WISE 3 color]{Star Formation and Gas Reservoir of UGC192. \textit{Left}: A WISE three-color image of UGC192, with 3.4 $\mmu$ (blue), 4.6 $\mmu$ (cyan/green) and 11.6 $\mmu$ (red). \textit{Middle}: \HI\ intensity distribution showing the disturbed morphology of UGC192 with a southern plume that has been suggested to be due to accretion of low column density primordial material (see \citealt{1998AJ....116.2363W,2019MNRAS.490.3365N}). Notice the spatial coincidence of the higher density \HI\ clumps with the bright SF regions in the 11.6 $\mmu$ emission. UGC192 is a star-bursting dwarf galaxy (a blue compact dwarf, BCD) in the local group. It has stellar mass $\mathrm{10^{8.6}M_\odot}$ much higher than its gas mass ($\mathrm{10^{7.4}M_\odot}$) and thus has a low gas fraction (0.06) that is similar to star forming spirals which place lower on the y-axis in the lower panels of Figure \ref{ml_mag}. Its absolute magnitude (stellar luminosity) is not atypical of the dwarfs in the sample (see left panels of Fig \ref{ml_mag}), but its surface brightness is exceptionally high (due to the starburst) and places it in the spiral-populated region in right panels of Figure \ref{ml_mag}. (see also Figure \ref{sfrmhi}). \textit{Right panel}: Combined \HI\ and three-color WISE image showing the extent of the obscured star formation with respect to the \HI\ disk. The green scale bar in the lower left corner shows an angular scale of 1 arcmin (0.2 kpc).}
\label{ic10}
\end{figure*}

UGC192 (IC10, a blue compact dwarf) is an outlier in all four plots. In the bottom panels, it shows much lower gas fraction ($\mathrm{M_{\HI}/M_*}$) than the rest of the dwarfs and is tending toward the spiral regime in luminosity and surface brightness. This is because (i) the quantities presented here are calculated within the W1 stellar disk, while UGC192 has been shown to have a significant amount of its \HI\ gas in an extended stream beyond the main disk \citep{1998AJ....116.2363W} and (ii) this galaxy is a starburst dwarf irregular galaxy exhibiting high star formation rates like the spirals and producing more energy than a regular dwarf, hence its high infrared luminosity (Figure \ref{ic10} shows a WISE three-color illustration of UGC192). On the other hand we have UGC12713 a dwarf spheroidal in a zone populated by late-type dwarfs. This shows that the apparent scaling of these quantities with morphology is not a cause but rather an effect of the underlying scaling relations amongst the quantities themselves and processes such as star formation. We find that galaxies with higher gas factions are intrinsically fainter in W1, i.e. they have lower stellar luminosity, than those with low gas fractions. \citet{1969AJ.....74..859R} suggested that the gas fraction may decrease towards earlier type spirals, which indeed shows in the lower panels of Figure \ref{ml_mag} where the early-type spirals, populate the low gas fraction part of the plots. The mass luminosity scaling relations in our sample are in agreement with those of \citet{2001A&A...370..765V} taken from the Ursa major galaxy cluster.

\begin{figure}
\includegraphics[scale=0.5]{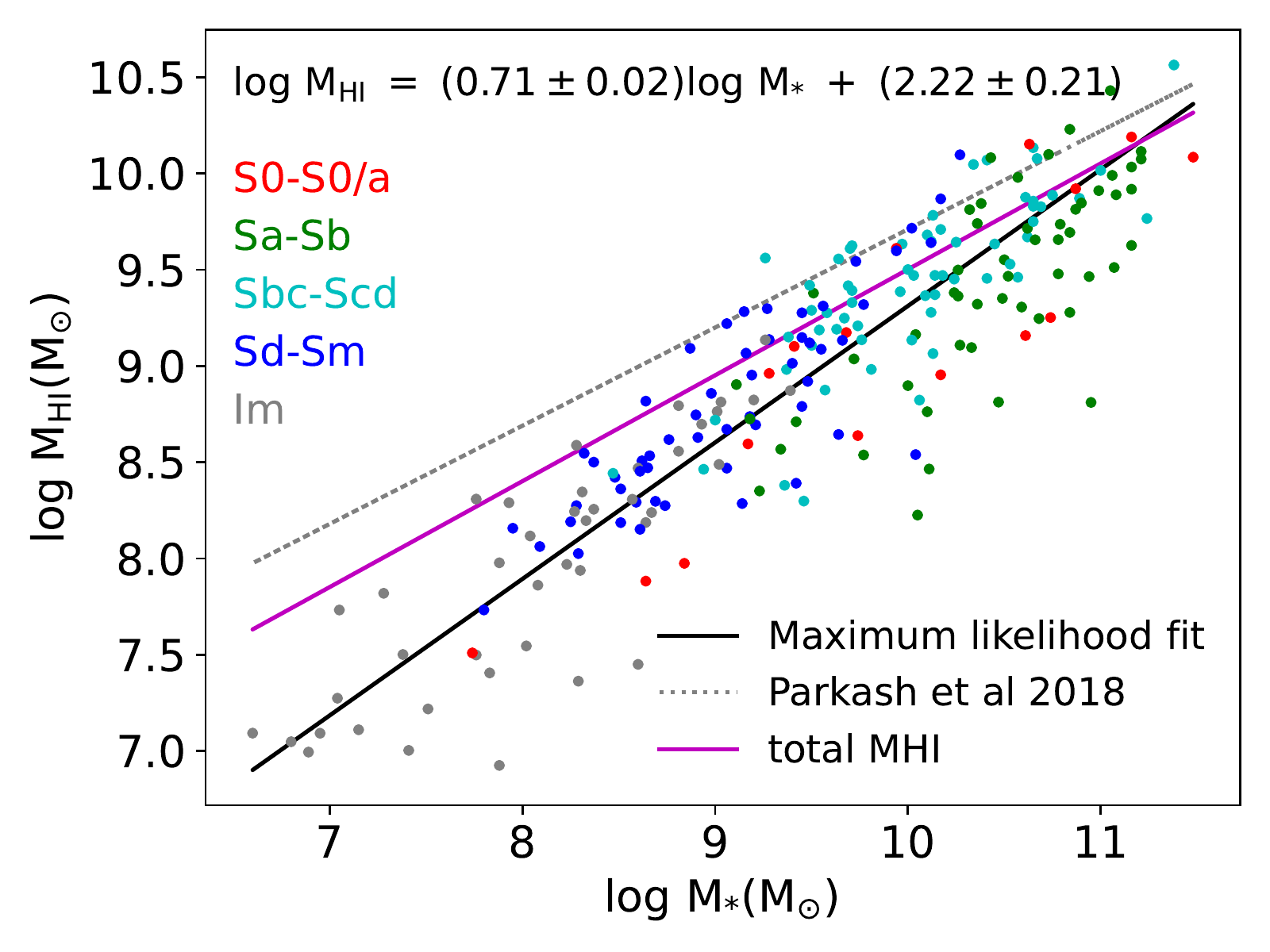}
\caption{Enclosed \HI\ mass versus stellar mass. The data points are color coded by the morphology of the galaxies such that the red represents lenticulars (S0-S0/a), green represents early type spirals (Sa-Sb), cyan represents Sbc - Scd, blue represents late-type spirals (Sd-Sm) and the gray are the late-type dwarfs. The solid black line indicates our best fit function to the data, with parameters shown in the upper left. Note that the \mhi\ used here is the \HI\ mass contained within the stellar disk (see Section \ref{stellarmasssection}). For comparison, a fit to the total \mhi\ (from the global profile) for our WHISP sample is shown in magenta and lies closer to the \citet{2018ApJ...864...40P} HICAT sample fit (gray line) which also used total \mhi .}
\label{mhi_mstar}
\end{figure}

%
The \mhi\ - $\mathrm{M_*}$ relation for our WHISP sample is plotted in Figure \ref{mhi_mstar}. In general, there is a relatively tight correlation whereby the \mhi\ increases with the $\mathrm{M_*}$, but the relation might be non-linear at the high-mass end. A maximum likelihood fit to the data yields a slope of $\mathrm{0.71\pm 0.02}$ (solid black line). At $\mathrm{\log\ M_*\ \sim\ 9}$, we observe the high mass Sa-Sb and S0 galaxies shifting below the relation, perhaps the beginning of an apparent migration off the relation by gas deficient early types. A flattening which depicts a lower dependence of the gas mass on the stellar mass at the high-mass end has been observed by studies such as \citet{2010MNRAS.403..683C,2012ApJ...756..113H,2015MNRAS.447.1610M,2016MNRAS.460.3419M}. Given that our sample includes very few S0's and no ellipticals, we, like \citet{2018ApJ...864...40P} and their HICAT-WISE study, are unable to see the actual flattening of the relation which occurs for the early types with low gas fractions (see \citealt{2010MNRAS.403..683C}). The salient difference in slopes between our study and \citet{2018ApJ...864...40P} is because we use the \HI\ mass enclosed within the stellar disk. For comparison, a fit using the total \mhi\ (from the global profile) for our WHISP sample is shown with the solid magenta line in Figure \ref{mhi_mstar}, and lies closer to the Parkash slope. The main factor is that the \HI\ disks of the gas rich spirals and dwarfs tend to be more extended than their optical disks \citep{2002A&A...390..829S,2013MNRAS.433..270W,2018MNRAS.478.1611K,2020ApJ...890...63W}. Given the low stellar masses of dwarfs, hence lower disk gravitational potential, the gas is less concentrated and the effect of using stellar disk enclosed \mhi\ is more pronounced in the dwarfs regime, leading to steepening of the \mhi\ - $\mathrm{M_*}$ slope at the low-mass end. In contrast, the fits tend to converge on the high-mass end, implying that most of the \mhi\ in the early types is contained within the stellar disks.

From the relations observed in this section we find that the global \HI\ properties of our galaxy sample are in agreement with other samples and scale as expected, although with a different slope because of the aperture matching between the \HI\ and the infrared. This means that our galaxy sample is a balanced representation of the general galaxy population in terms of the \HI\ content and stellar luminosity.

\newpage
\subsection{\textit{The Star formation main sequence relation}}\label{section:sfms}
The correlation between star formation rate and stellar mass, known as the star formation main sequence (SFMS), is a well established scaling relation in galaxies  \citep{2004MNRAS.351.1151B,2007ApJ...660L..47N,2007ApJ...670..156D,2010A&A...518L..25R,2011ApJ...742...96W}  with a remarkably tight scatter of ~0.2 dex,  and has been shown to hold over all redshifts up to z=6 \citep{2014ApJS..214...15S}. 
This relation depicts the star forming histories (SFH) of galaxies in the universe since it compares the instantaneous star formation to past star formation. It defines the rate at which galaxies build their mass, with the more massive galaxies assembling their mass earlier on in cosmic history. The SFMS is therefore an important constraint on the mass assembly histories of galaxies and is widely applied in models of galaxy formation and evolution (e.g., \citealt{2010ApJ...721..193P,2012ApJ...745..149L,2013ApJ...770...57B,2015MNRAS.451.2663H,2015MNRAS.447.3548S}). The scatter in the relation has been observed to increase at the high mass end, a feature attributed to the widely varying SFHs among massive galaxies including minor and satellite merger events.

\begin{figure*}
\includegraphics[scale=0.35]{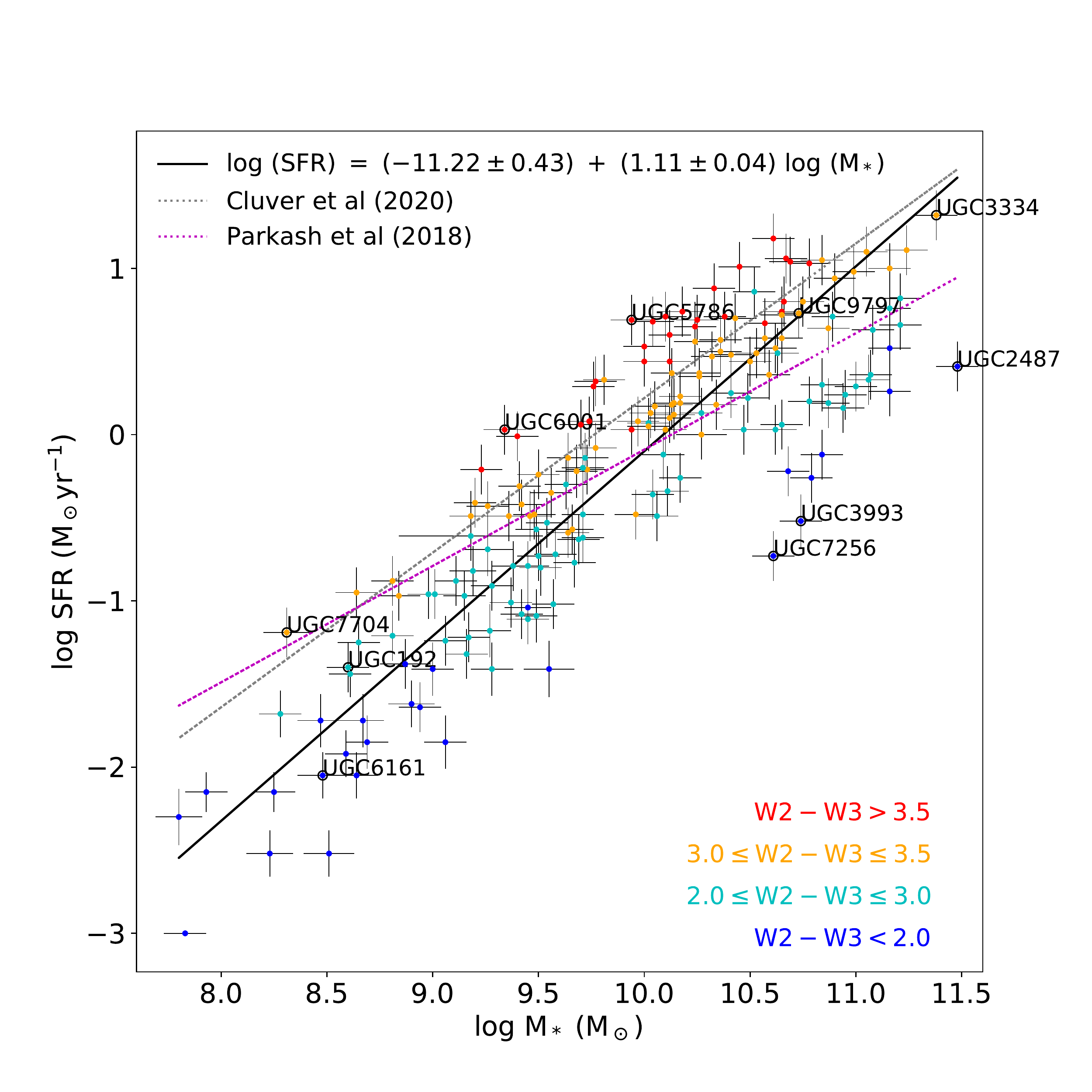}
\caption[Star formation main sequence ($SFR-M_*$)]{Relation between SFR and stellar mass for 180 galaxies detected in W3. The plot is color coded by the W2-W3 color where blue, cyan, orange and red represent W2-W3 $\leq2$, $\leq3.0$, $\leq3.5$ and, $\geq3.5$ respectively. The W2-W3 is an indicator of the dust content but also follows the morphologies such that low star forming early-types have bluer colors while the high star forming intermediate spirals have redder colors (we have followed the color-code convention of \citealt{2017ApJ...850...68C}). The solid black line is a maximum likelihood fit to the data. For a given mass range, the main sequence line marks a separation between early-types and late-types.}
\label{sfmainsequence}
\end{figure*}

Figure \ref{sfmainsequence} illustrates our main sequence relation for a sub-sample of 180 galaxies which are detected in the W3 band. The data are color coded by the W2-W3 color where blue, cyan, orange and red represent W2-W3 $\leq2$, $\leq3.0$, $\leq3.5$ and, $\geq3.5$ respectively. The W2-W3 is an indicator of the dust content but also roughly follows the morphologies such that low star forming early-types have bluer colors in the mid-IR bands while the high star forming intermediate spirals have redder colors (following the color-code convention of \citealt{2017ApJ...850...68C}). 
Our sample galaxies follow a well defined main sequence with a scatter of 0.4 dex. In agreement with previous studies (e.g \citealt{2015ApJ...808L..49G,2015A&A...579A...2I,2017ApJ...836..182J,2017ApJ...850...68C}), there is an increase in scatter at the high mass end partly due to the prevalence of gas exhaustion in high mass systems \citet{2007ApJ...660L..47N}. Passive evolution as well as processes such as mergers, quenching or gas stripping lead to declining SFRs, eventually causing the affected galaxy to migrate from the main sequence or its current SFR being decoupled from its past SFR \citep{2015A&A...579A...2I,2017ApJ...836..182J,2020arXiv200607535C}. 

Having decomposed their sample galaxies into disks and bulges, \citet{2015ApJ...808L..49G} suggest that the increased variations in the SFHs of massive galaxies are due to the prevalence of central dense structures such as bars and bulges in massive systems. A bulge that is not star forming contributes to the total stellar mass of a galaxy but not to the total star formation, yielding a specific star formation rate ($sSFR\ =\ SFR/M_*$) lower than that for a disk dominated system. The varying masses of the bulges affect the derived sSFR to different extents hence we see scatter in the relation, leaving the disk dominated systems to have more similar SFHs and hence a tighter main sequence relation. This may well be the reason for the clear segregation seen between the different morphological types in Figure \ref{sfmainsequence}, suggesting different SF main sequences for the different galaxy types. 
\citet{2020arXiv200607535C} have demonstrated the sensitivity of the SFMS relation to sample selection criteria such that star forming samples yield steeper relations than samples which include galaxies that have turned over at the high mass end. Indeed their sample which excludes the low mass - low SFR population ($\mathrm{M_*\ <\ 10^{9.75}, W2-W3\ \leq\ 3}$) has a shallower SFMS ($\mathrm{slope\ =\ 0.93}$) than our sample which includes $\mathrm{M_*\ <\ 10^{9}}$ and $\mathrm{W2-W3\ \leq\ 2}$. 
Likewise, we compare our relation to the SFMS relation of \citet{2018ApJ...864...40P}. Their SFMS was modeled using high stellar-mass ($\mathrm{M_* \geq 10^{10.5}\ M_\odot}$) spirals, close to the turnover at high-mass, resulting in an even shallower slope of 0.7. 
\begin{figure*}
\centering
\includegraphics[scale=0.35]{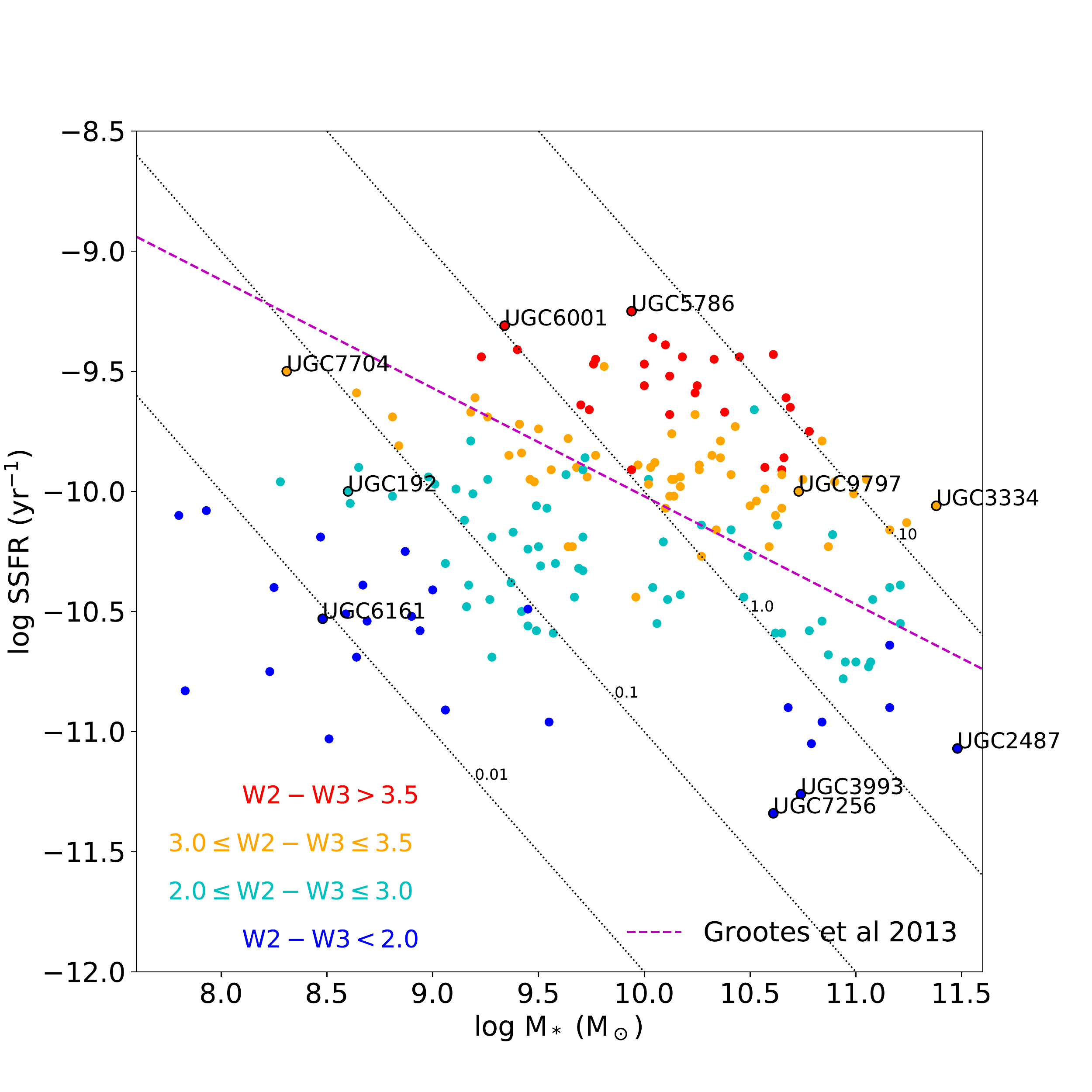}
\caption[Star formation main sequence ($sSFR-M_*$) ]{Relation between the specific SFR ($sSFR\ =\ SFR/M_*$) and stellar mass. The plot is color coded by the W2-W3 color where blue, cyan, orange and red represent W2-W3 $\leq2$, $\leq3.0$, $\leq3.5$ and, $\geq3.5$ respectively. The black dashed diagonal lines represent lines of constant SFR (0.01, 0.1, 1, and 10 $\mathrm{M_\odot yr^{-1}}$). The dashed magenta line is the relation of \citet{2013ApJ...766...59G} whose sample of spirals had $\log M_* \geq 9.6$.}
\label{ssfr}
\end{figure*}

\begin{table*}
\caption[Selected properties]{\textbf{Specific properties of galaxies labeled in Figures \ref{sfmainsequence} - \ref{sfrmhi}}: (1) - Name. (2) - stellar mass. (3) - Atomic hydrogen mass. (4) - Star formation rate as measured from W3 flux calibration. Note that the SFR is integrated over the stellar disk. (5) - Specific star formation rate. (6) - $\mathrm{W2-W3}$ color index. (7) - Gas fraction, that is, the ratio of \HI\ mass to the stellar mass, both quantities summed up inside the stellar disk. (8) - Morphology.}
\begin{tabular}{cS[table-format=3.2]S[table-format=3.2]S[table-format=3.2]S[table-format=3.2]S[table-format=3.3]S[table-format=3.3]c}
\hline \hline
Name & {$\log (M_*)$} & {$\log (M_{\HI})$} & {$\log (SFR)$} & {$\log (sSFR)$} & {W2-W3} & {$\frac{M_{\HI}}{M_*}$} & Morph \\
 & {$M_\odot$} & {$M_\odot$} & {$M_\odot yr^{-1}$} & {$yr^{-1}$} & {mag} & {} &  \\
(1) & {(2)} & {(3)} & {(4)} & {(5)} & {(6)} & {(7)} & (8) \\
\hline
UGC192 & 8.6 & 7.45 & -1.4 & -10.0 & 2.75 & 0.07 & Im \\
UGC2487 & 11.48 & 10.08 & 0.41 & -11.07 & 1.83 & 0.04 & S0 \\
UGC3334 & 11.38 & 10.56 & 1.32 & -10.06 & 3.39 & 0.15 & Sc \\
UGC3993 & 10.74 & 9.25 & -0.52 & -11.26 & 1.35 & 0.03 & S0 \\
UGC5786 & 9.94 & 9.61 & 0.69 & -9.25 & 4.26 & 0.47 & pec \\
UGC6001 & 9.34 & 8.57 & 0.03 & -9.31 & 4.03 & 0.17 & Sa \\
UGC6161 & 8.48 & 8.42 & -2.05 & -10.53 & 1.8 & 0.88 & Sdm \\
UGC6713 & 8.51 & 8.36 & -2.52 & -11.03 & 0.82 & 0.71 & Sm \\
UGC7232 & 7.83 & 7.41 & -3.0 & -10.83 & 1.26 & 0.38 & Im \\
UGC7256 & 10.61 & 9.16 & -0.73 & -11.34 & 1.15 & 0.04 & S0 \\
UGC7704 & 8.31 & 8.35 & -1.19 & -9.5 & 3.25 & 1.09 & Im \\
UGC9797 & 10.73 & 10.1 & 0.73 & -10.0 & 3.48 & 0.23 & Sb \\
\hline
\end{tabular}
\label{specialprops}
\end{table*}

In Figure \ref{ssfr}, we plot the sSFR against stellar mass and find a mostly flat trend which turns downward, tending to a negative slope, for the high mass galaxies. This trend, which was also found by \citet{2017ApJ...836..182J} and \citet{2020arXiv200607535C}, shows that even though the low mass galaxies have lower SFRs than the high mass systems, the former are still in an active stage of building their disks while the latter are progressing toward passive or quiescent star forming states \citep{2017ApJ...836..182J}. For example UGC7256 (NGC4203) is a lenticular galaxy with ongoing SFR but given its stellar disk of $\mathrm{\sim 10^{10.6}\ M_\odot}$ (and stellar mass to light ratio = 0.73), it has a low sSFR and is no longer actively building its disk. This is also the case for UGC3993 and UGC2487 (NGC1167). These three also have low gas fractions ($\leq$ 0.04). On the other hand UGC3334 (NGC1961) has a considerable bulge to total disk ratio (0.71), has one of the highest stellar masses in the sample ($\mathrm{\sim 10^{11.4}\ M_\odot}$), but is also dusty (W2-W3 = 3.25) with a higher gas fraction (0.15) and sSFR ($\mathrm{\sim 10^{-10.1}\ yr^{-1}}$). Note that in high mass systems there can be cases of enhanced star formation even after the galaxy has assembled its mass, for example in the case of merger events. Such events will cause the affected galaxy to migrate upwards and as the enhanced SF period phases out it will migrate again down towards passive SF. We find, as did \citet{2020arXiv200607535C}, that it is the more dusty systems (large W2-W3 colors) which have the highest sSFR. For example, UGC5786 (NGC3310) and UGC 6001 (NGC3442) have the highest sSFR and highest dust content, ($\mathrm{W2-W3\ \geq\ 4}$).  \\
We have over-plotted the sSFR-M$_*$ relation of \citet{2013ApJ...766...59G} in Figure \ref{ssfr}. Their sample-selection was based on NUV detection which is more more sensitive to relatively dust-free SF systems, notably dwarf systems, and hence the high sSFR for the low mass end (see \citealt{2017ApJ...836..182J}).

Note that there are few late-type dwarfs in our SFMS sub-sample because most of them have very low dust contents and lie below the W3 detection threshold. The W1 band is very sensitive and well able to detect these dwarfs and their stellar mass, but this is not the case for W3. 
Also note; Table \ref{specialprops} lists specific properties of the galaxies labelled in Figures \ref{dhi_mhi} - \ref{sfrmhi}, for reference with the plots.

\subsection{\textit{Star formation rate vs \HI\ mass}}\label{section:sfrmhi}
The disruption of the atomic gas leads to suppression of star formation because atomic gas is needed to cool and form the molecular clouds \citep{2008A&A...490..571F,2015ApJ...805..145E} from which the stars form.

\begin{figure*}
\centering
\includegraphics[scale=0.33]{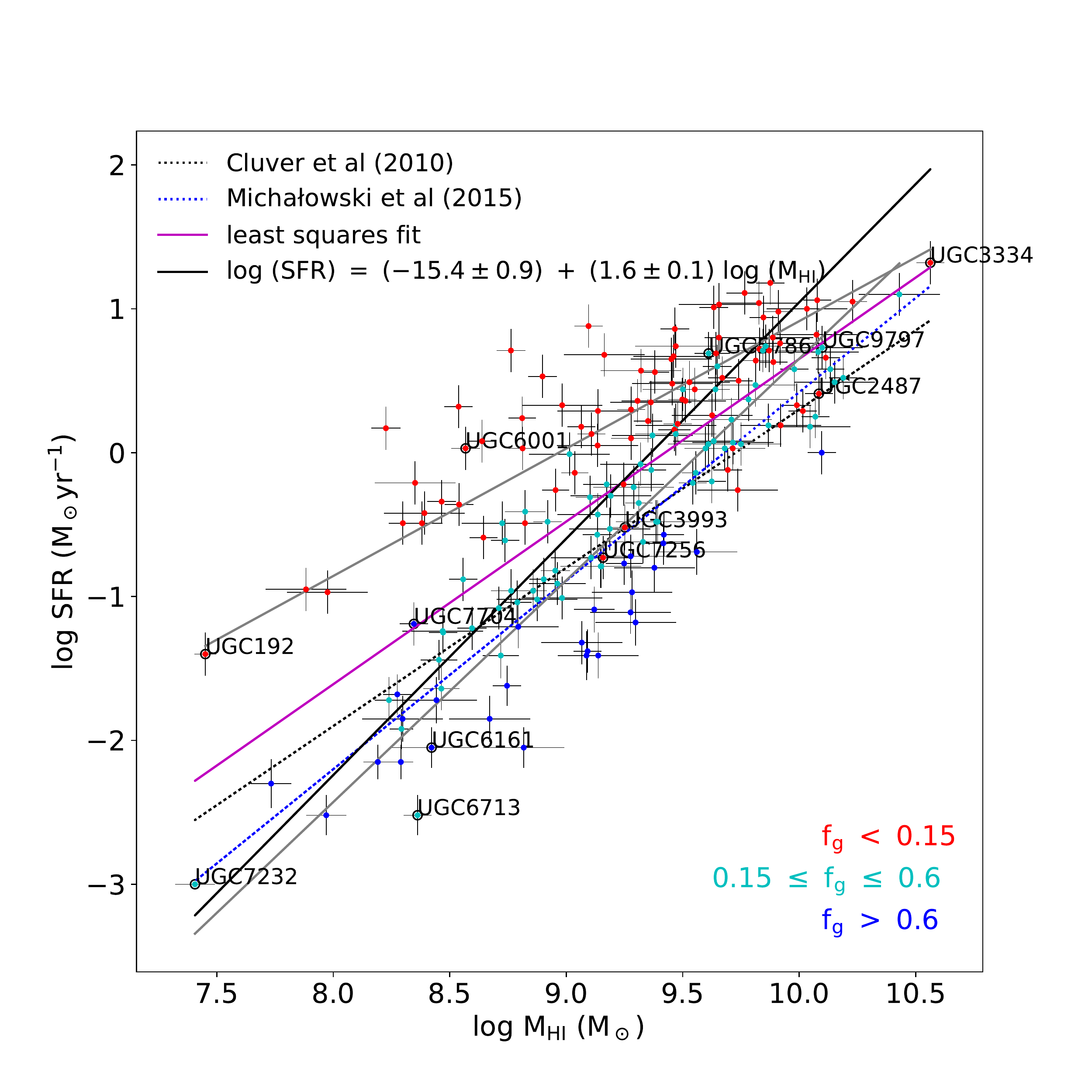}
\includegraphics[scale=0.33]{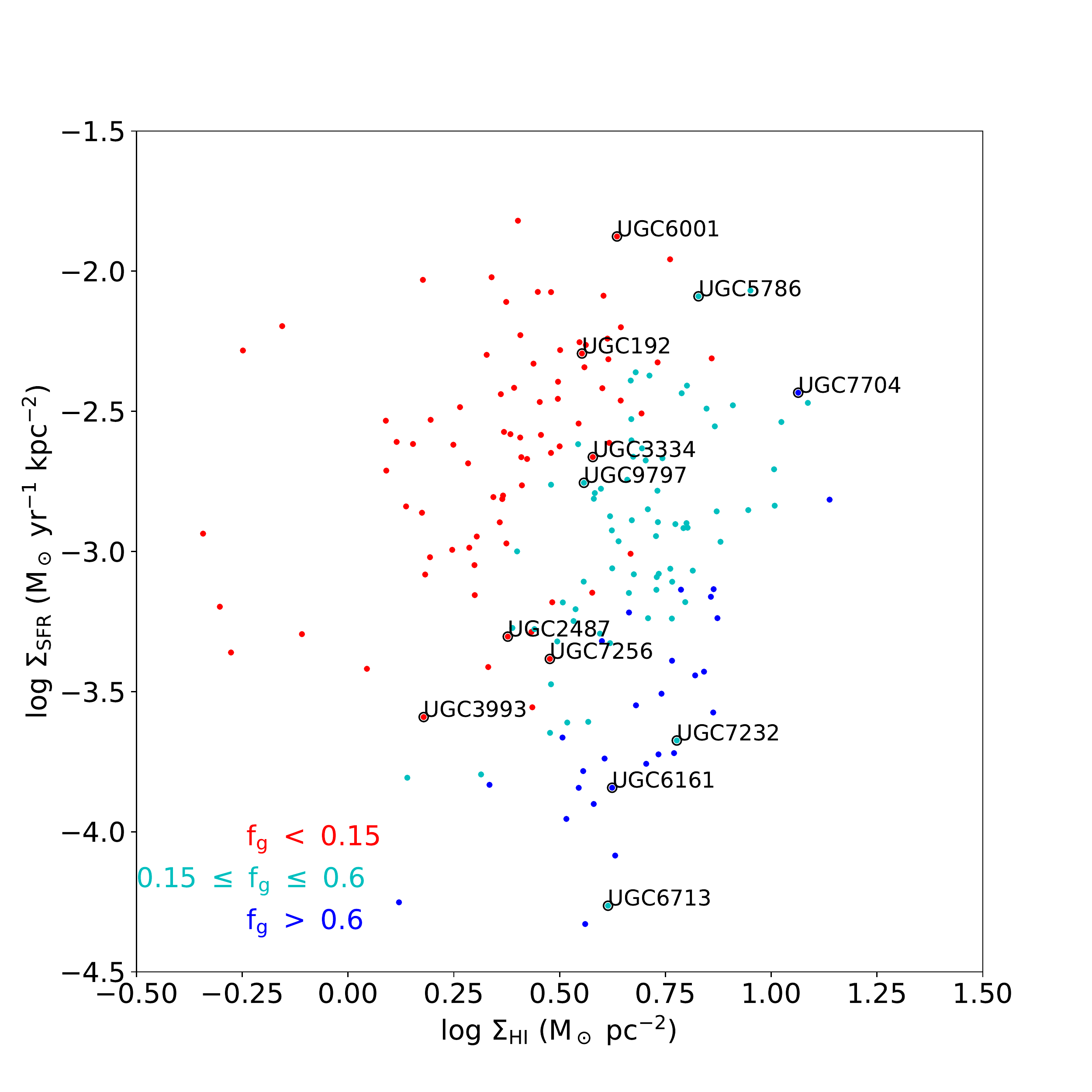}
\caption[SFR vs Atomic gas mass ]{\textit{Left panel}: Star formation rate versus atomic gas mass. The \HI\ masses were derived from gas enclosed within the stellar disk defined by the 1$\sigma$ isophote in the W1 images ($\sim$23 \marc in Vega units). The points are color-coded by gas fraction ($\mathrm{f_g = MHI/M_*}$) such that red represents $\mathrm{f_g \leq 0.15}$, cyan for $\mathrm{0.15 \leq f_g \leq 0.6}$ and blue for $\mathrm{f_g > 0.6}$. The color code highlights a `green valley' and apparent red and blue sequences (fit by the two thin gray lines - discussed in the text) with a few outliers. Our maximum likelihood best fit (solid black line) is shown together with the lines of \citet{2010ApJ...725.1550C} and \citet{2015A&A...582A..78M}. \textit{Right panel}: Mean star formation rate surface density versus \HI\ surface density for the sub-sample. The surface densities were averaged over the entire stellar disk defined by the 1$\sigma$ isophote in the W1 images. The plot is color coded by gas fraction ($\mathrm{f_g = MHI/M_*}$) such that red represents $\mathrm{f_g \leq 0.15}$, cyan for $\mathrm{0.15 \leq f_g \leq 0.6}$ and blue for $\mathrm{f_g > 0.6}$. There is no correlation observed between the \sighi\ and \sigsfr\ at these global scales.}
\label{sfrmhi}
\end{figure*}

In Figure \ref{sfrmhi} (left panel), we plot the SFR as a function of the atomic gas mass enclosed within the stellar disk. The plot is color-coded by gas fraction ($\mathrm{f_g = M_{\HI}/M_*}$) such that red represents $\mathrm{f_g \leq 0.15}$, cyan $\mathrm{0.15 \leq f_g \leq 0.6}$ and blue $\mathrm{f_g > 0.6}$. 
%
%
The galaxies with low gas fractions are mostly early type spirals such as UGC2487 (S0), UGC3993 (S0) and UGC6001 (Sa). Right away, it is clear that the relationship between SFR and $\mathrm{M_{\HI}}$ holds at different $\mathrm{f_g}$. There is a general trend that at any given gas mass, the galaxies with low $\mathrm{f_g}$ have higher SFR, a trend that remains even when the parameters are normalized by surface area (see right panel of Figure \ref{sfrmhi}). 

Overlaid on the data in Figure \ref{sfrmhi} (left panel) are; a least squares fit with slope 1.1$\pm$0.1, a best fit line (maximum likelihood\footnote{http://hyperfit.icrar.org/, \citet{2015PASA...32...33R}}) with a slope of 1.6$\pm$0.1, as well as the relations of \citet{2010ApJ...725.1550C} for galaxies in the Spitzer Infrared Nearby Galaxies survey (SINGS\footnote{\label{ss}http://irsa.ipac.caltech.edu/data/SPITZER/SINGS/}) and \citet{2015A&A...582A..78M}. These two relations have slopes 1.1 and 1.3 respectively, in agreement within the errors, with our least squares fit which only minimizes the square residuals overall. However, the maximum likelihood fit highlights a feature in the data where the high mass end is biased by galaxies with low $\mathrm{f_g}$ while the low mass end is biased by galaxies with high $\mathrm{f_g}$. These two groups separately appear to have different relations. To investigate this, we binned the data by gas fraction into two groups separated at $\mathrm{f_g\ =\ 0.15}$ and applied maximum likelihood fits to each. The thin gray lines are the result showing a shallow slope (0.9$\pm$0.1) for $\mathrm{f_g\ \leq\ 0.15}$ (the red color code which are early-type spirals), and a steeper slope (1.5$\pm$0.1) for $\mathrm{f_g\ >\ 0.15}$ (blue and cyan, which are intermediate and late-type). The early types, even though having higher total atomic gas mass, have already assembled the bulk of their stellar mass and hence have lower gas fractions. Likewise their SFR's do not increase significantly with mass, because they are moving into a more quiescent stage of star formation (see Section \ref{section:sfms}), hence resulting in a shallow relation. On the other hand, the later type spirals are still actively building their disks and still have high gas fractions resulting in a steeper difference between SFR for different $\mathrm{M_{\HI}}$.

In the right panel of Figure \ref{sfrmhi}, we plot the surface densities of SFR and $\mathrm{M_{\HI}}$. Note that here we refer to the integrated \HI\ surface density, obtained from dividing the $\mathrm{M_{\HI}}$ enclosed in the stellar disk (as defined in Section \ref{stellarmasssection}) by the de-projected area of the stellar disk. We do not trace any correlation between the two quantities, due to the limited range of \HI\ surface densities (one magnitude) spanned by our sample. This limitation partly arises from the tight mass-size relation of spiral galaxies (see Section \ref{hi_diams_sec}) whose consistent slope of $\sim 2$ for a wide variety of samples \citep{2016MNRAS.460.2143W} implies a roughly constant characteristic average \HI\ surface density for all star forming galaxies \citep{2001A&A...370..765V,2005A&A...442..137N,2011PhDT.......327M}. The SFR, on the other hand, is not regulated by disk size (or projected surface area) so the \sigsfr\ spans a wider range of three orders of magnitude. Note that our entire sample falls on the steep part of the relation of \citet{2008AJ....136.2846B} (See their Figure 10).

A study of the resolved disks rather than global averages would give a fair comparison between the surface densities of atomic hydrogen and SFR. 
In Paper II, we carry out a resolved study of the gas density threshold for star formation in light of two models of disk stability, where we investigate the necessary critical density/stability parameter for the onset of star formation. 

\section{Summary and Conclusions}\label{theconclusion}
Having derived new \HI\ intensity maps for the WHISP survey, we have measured global relations between \HI\ and stellar disk properties using WHISP 21cm line observations and WISE 3.4 $\mmu$ imaging for 228 galaxies. We have also studied the star forming main sequence (SFR vs M$_*$) for a sub-sample of 180 galaxies detected in the WISE 12 $\mmu$ band, as well as the relationship between SFR and \mhi . We find the following:\\
\begin{itemize}
\item[1.] Our sample of 228 galaxies fall on the \HI\ mass-size relation with slope 1.95$\pm$0.03, and a characteristic mean \HI\ surface density, both in agreement with the literature on studies of complete samples.
\item[2.] Correlations of the \mhi\ and \HI\ mass-to-light ratio ($\mathrm{M_{\HI}/L}$) with stellar luminosity and surface brightness ($\mu_{W1}$) have lower scatter when considered for \mhi\ enclosed within the stellar disk than when total integrated masses are used. We obtained vertical scatter of 0.17 for the relations with stellar luminosity while the relations for $\mathrm{\mu_{W1}}$ have a scatter of 0.37 for \mhi\ and 0.21 for $\mathrm{M_{\HI}/L}$. Thus the relations for stellar luminosity are tighter than those for stellar surface brightness.
\item[3.] The general trends in our relations are the same as in other studies which use integrated properties over the entire \HI\ disk. The \mhi\ is negatively correlated with the stellar luminosity and $\mathrm{\mu_{W1}}$ (R=0.9 and R=0.7 respectively) while $\mathrm{M_{\HI}/L}$ is positively correlated with the same (R=0.76 and R=0.78). Note that we obtain a tight relation for $\mathrm{M_{\HI}}$ vs $\mathrm{M_{W1}}$ as a result of comparing spatially co-located quantities in the stellar disk.
\item[4.] The early-type spiral galaxies tend to have higher stellar luminosity and higher \mhi , but lower $\mathrm{M_{\HI}/L}$ than late-type spirals and dwarfs.
%
\item[5.] Our sample follows a SF main sequence of slope 1.1 with a vertical scatter of 0.4. There is a segregation between the galaxies with higher dust content (which typically have higher SFR at any given constant stellar mass) above the main sequence and the galaxies with lower dust content (lower SFR) below the main sequence. 
\item[6.] There is a characteristic migration of the high M$_*$ galaxies off the main sequence towards lower SFR, which may be due to quenching, gas stripping or mergers the end result of which is passive evolution.
\item[7.] We find a correlation between SFR and \mhi\ in agreement with the literature, albeit with a large scatter of 0.6. The SFR-\mhi\ relation appears to be driven by the gas fraction and stellar mass of the galaxies.
\item[8.] We find no correlation between the global average surface densities of SFR and atomic gas because of the small range in magnitude spanned by \sighi . This is, atleast in part, due to the \HI\ mass-size scaling relation which results in a roughly constant characteristic average surface density for \HI\ disks.  
\end{itemize}

\section*{Acknowledgments}
We sincerely thank the anonymous reviewer for insightful comments that greatly improved the quality of this paper. Great thanks to Thijs van der Hulst, Danielle Lucero, Michelle Cluver, Natasha Maddox, Janine van Eymeren, Rob Swaters, Adam Leroy, Amidou Sorgho, Maria Kapala, Mpati Ramatsoku, Claude Carignan, Erwin de Blok, and Ren\'{e}e Kraan-Korteweg for useful discussions around this work. \\
EN acknowledges that this research was funded by the South African Research Chairs Initiative (SARCHI) of the Department of Science and Technology (DST) and the National Research Foundation (NRF).\\
ECE acknowledges support from the South African Radio Astronomy Observatory, which is a facility of the National Research Foundation, an agency of the Department of Science and Technology.\\
THJ acknowledges support from the National Research Foundation (South Africa)

\section*{Data Availability}
The data products (\HI\ distribution maps and velocity fields) generated in this research will be made available upon reasonable request to the corresponding author.

\bibliographystyle{mnras}
\bibliography{WISE_WHISP}

\appendix
%
%
%

\section{Sample selection based on infrared photometry}
\begin{center}
\begin{figure*}
\subfloat[]{\includegraphics[scale=0.35]{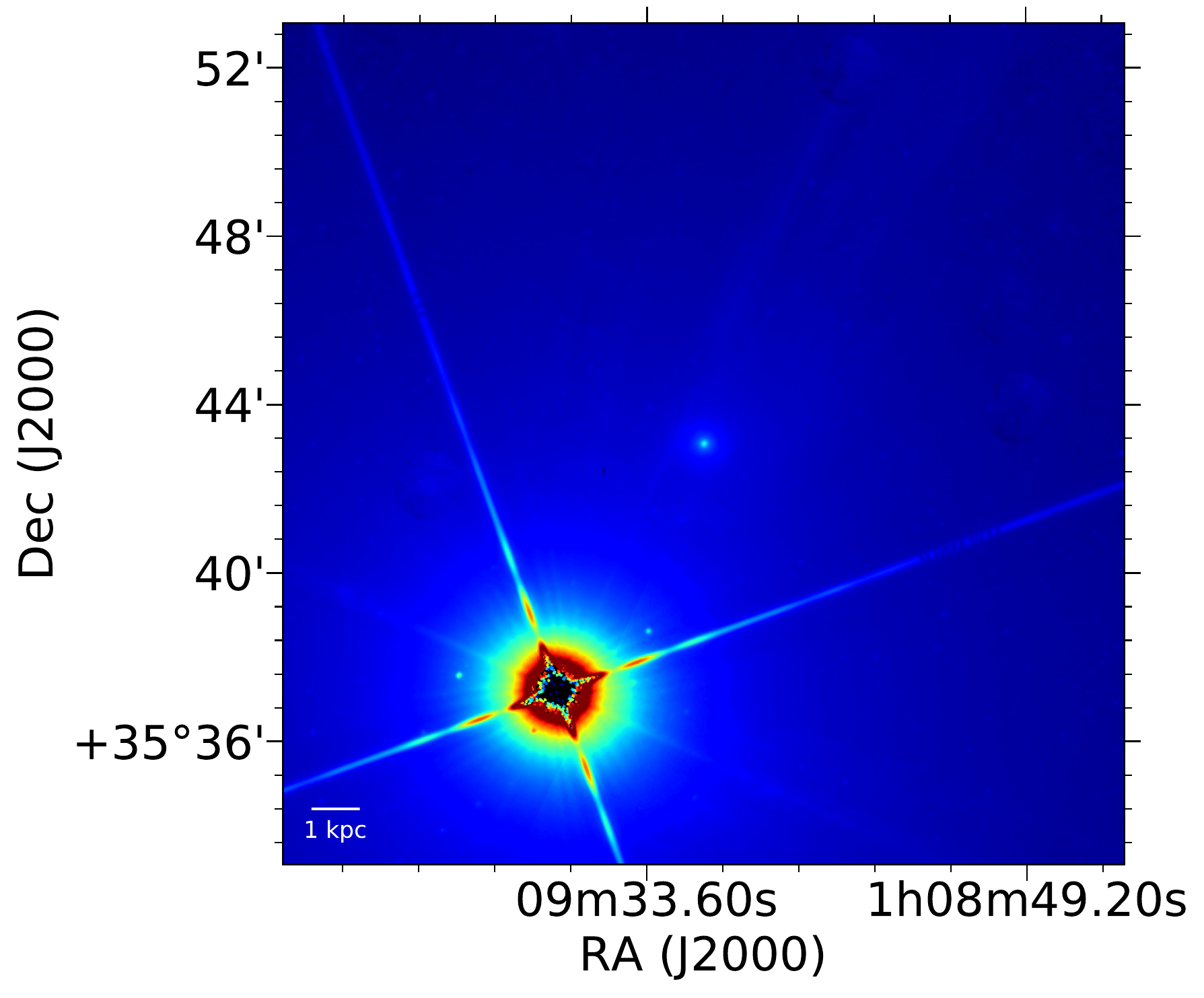}}
\quad
\subfloat[]{\includegraphics[scale=0.35]{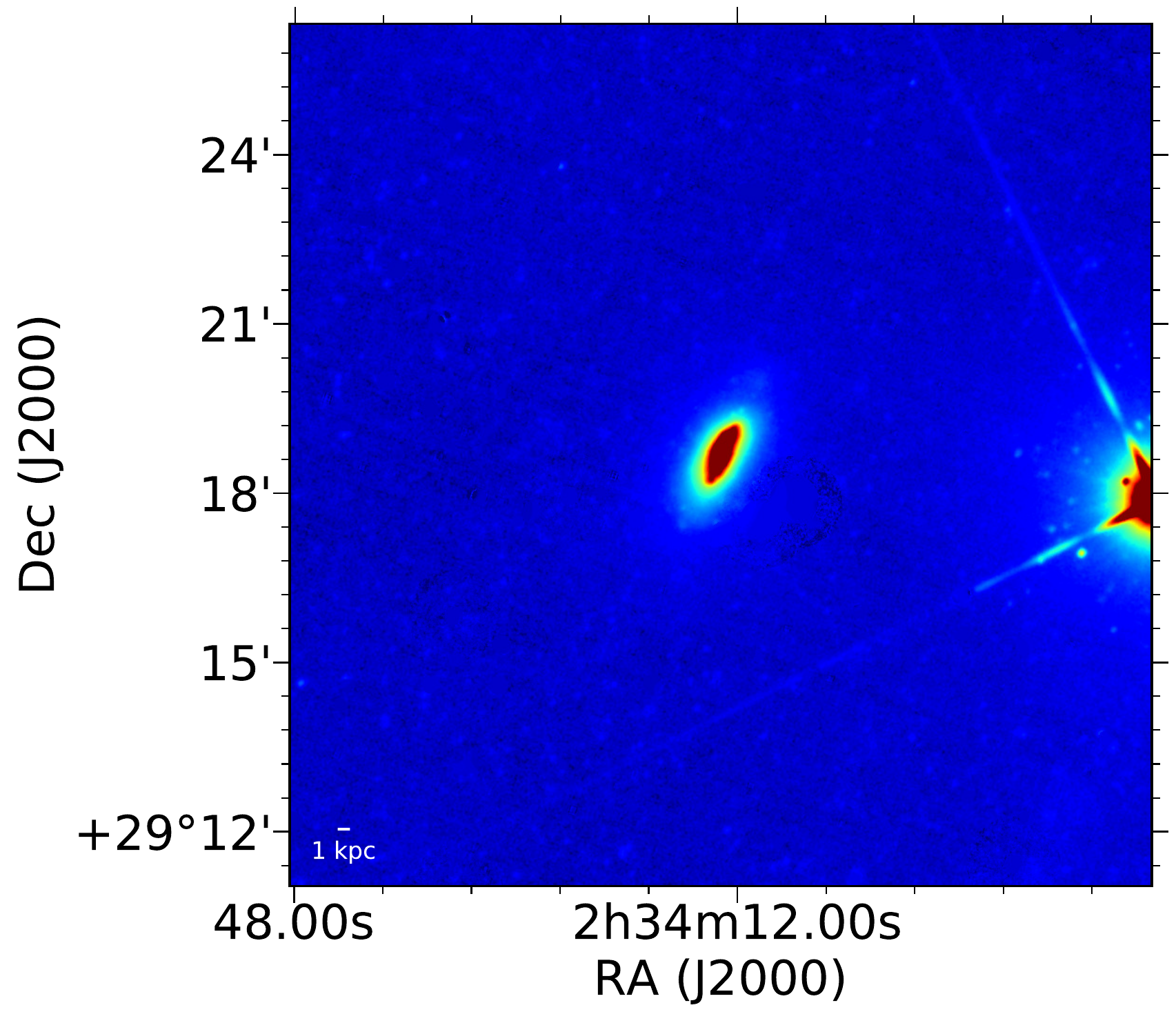}} \\
\quad
\subfloat[]{\includegraphics[scale=0.35]{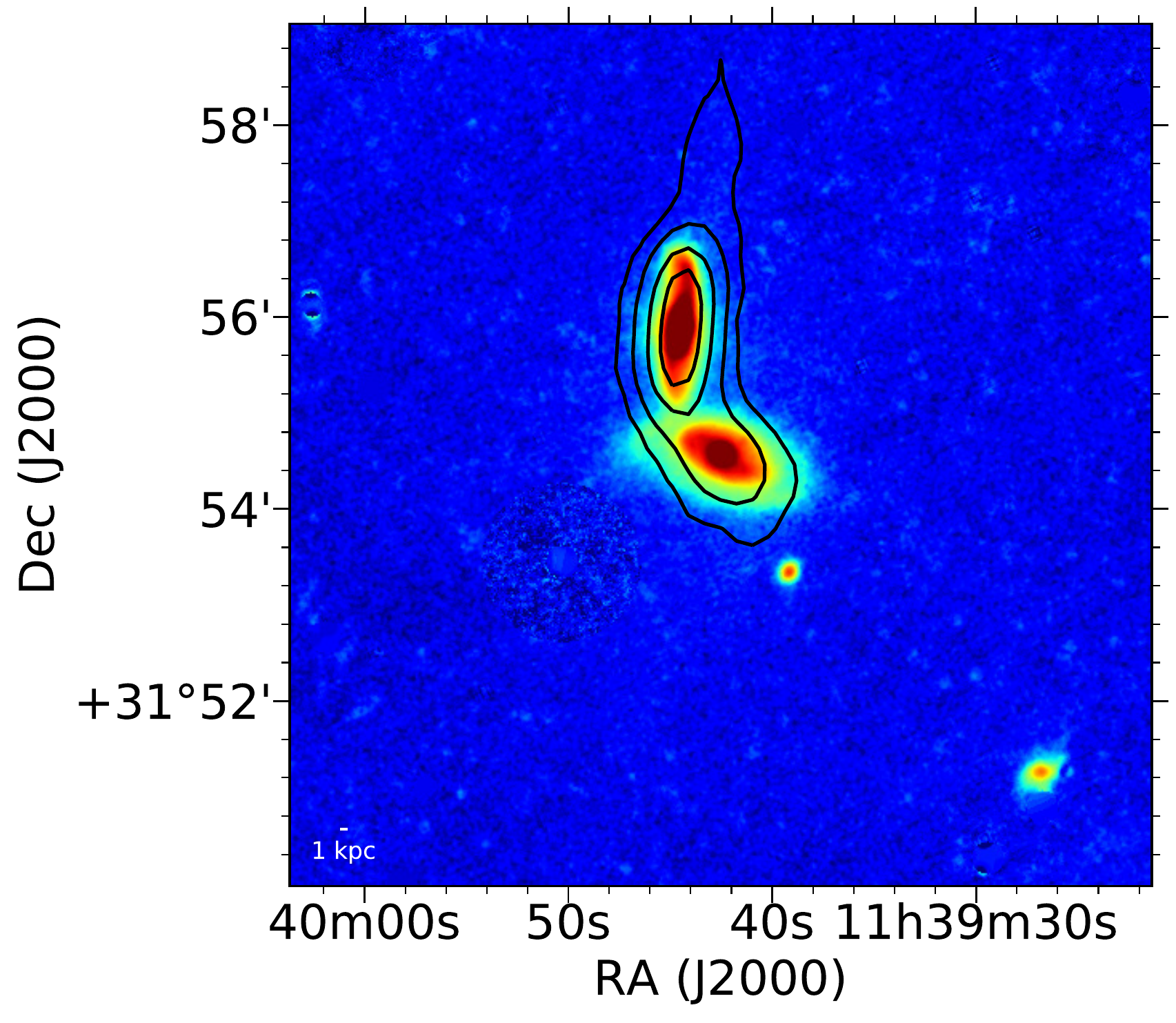}}
\quad
\subfloat[]{\includegraphics[scale=0.35]{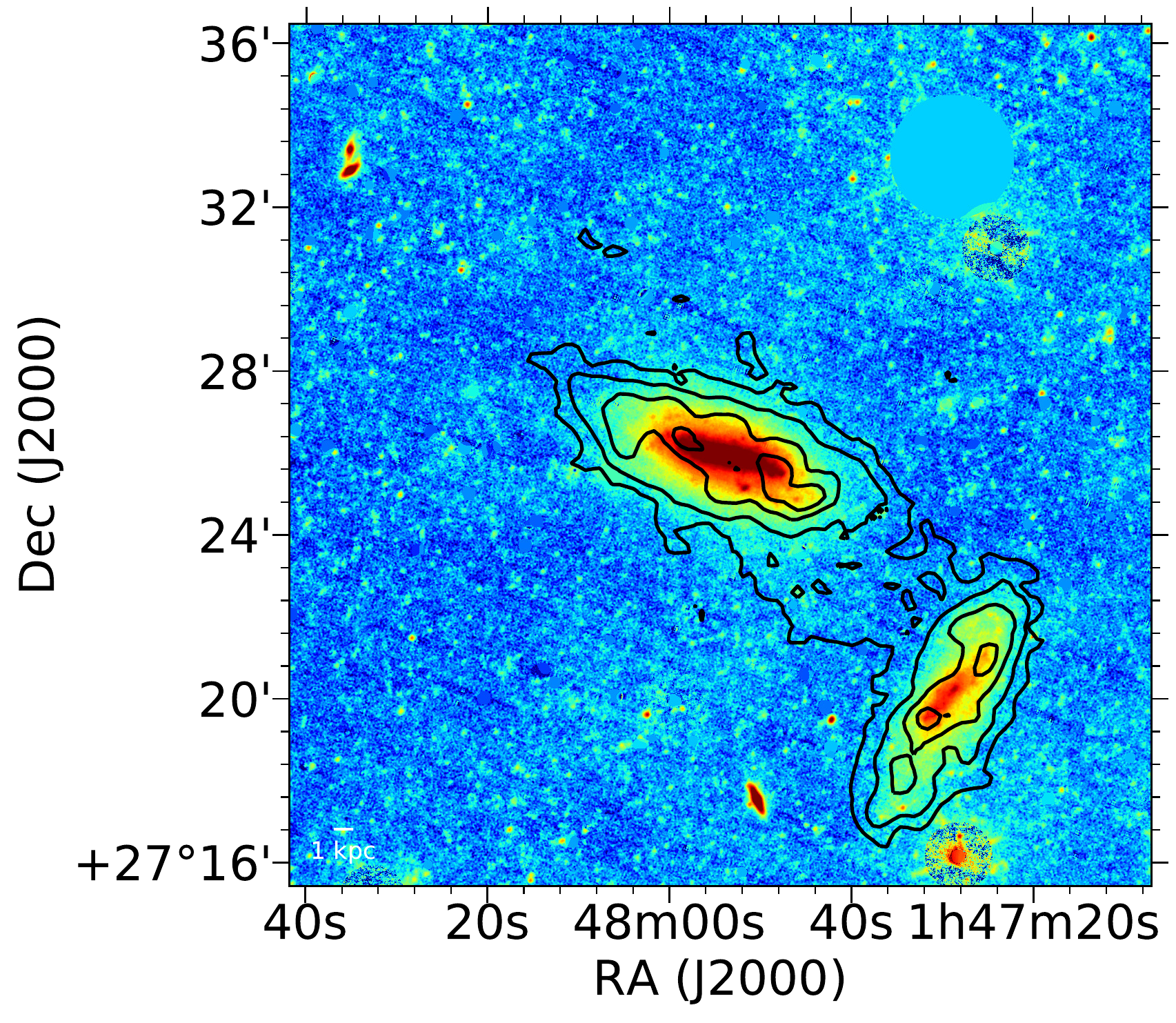}}
\caption[Photometry check]{Top panels - WISE W1 images with foreground stars; Galaxies where the image flux was dominated by a foreground star (a) were dropped from the sample, unlike images where a bright foreground star did not dominate the flux (b). Bottom panels - W1 images of interacting galaxies. Black contours trace the \HI\ distribution at levels of 0.032 $\mathrm{Jy\ \kms}$, 0.081 $\mathrm{Jy\ \kms}$ , 0.16 $\mathrm{Jy\ \kms}$ and 0.24 $\mathrm{Jy\ \kms}$; Wherever the interaction was close enough to be picked up in the infrared, such targets were dropped from the sample. Panel (c) shows one such example (UGC6621/UGC6623), while panel (d) shows a case where the infrared stellar disks are well separated (UGC1256/UGC1249).}
\label{bad_phot}
\end{figure*}
\end{center}

When choosing our sample from the WHISP survey, we considered the photometry in the corresponding WISE 3.4 $\mmu$ (W1) images. Galaxies with bright foreground stars whose flux dominated the W1 flux as well as galaxies with interacting stellar disks were rejected. Figure \ref{bad_phot} shows examples of accepted and rejected cases.
%
\section{Adopted distances}
The distances used in this study were obtained from two sources: the Cosmic Flows catalog (CF2) \citep{2013AJ....146...86T} via the Extragalactic Distance Database (EDD)\footnote{\label{*}http://edd.ifa.hawaii.edu/} \citep{2009AJ....138..323T} and the NASA/IPAC Extragalactic Database (NED)\footnote{https://ned.ipac.caltech.edu/}. The NED distances are estimates from redshift-derived radial velocities ($\mathrm{V = cz}$) corrected for peculiar flows due to the influence of the Virgo cluster, the Great Attractor and the Shapley supercluster \citep{2000ApJ...529..786M}. The EDD distances are based on different surveys, namely, The 2Micron All Sky Survey Extended Source Catalog (2MASX, \citealt{2000AJ....119.2498J}), The Catalog of Neigbouring Galaxies \citep{2004AJ....127.2031K}, and the 2MASS V8k catalog \citep{2005ASPC..329..135H}, as well as independent distance derivations from archival data by \citet{2009AJ....138..323T}.These sources had varying methods of deriving their distances, and hence the EDD distances are weighted averages of several methods; the Cepheid period-luminosity relation (C), Tip of the Red Giant Branch method (TRGB), Supernovae type 1a (SNIa), Tully-Fisher relation (TFR) method and the Fundamental plane (FP) method. The Cepheids method and FP method get the highest and lowest weights respectively. When obtaining distances from the EDD, we took only distances with estimates from the first four methods. The EDD database assigns a 20\% uncertainty to the distance estimates. For galaxies that were not listed on EDD or had only FP-based estimates, distances were obtained from the NED Mould flow estimates. Since the distance estimates given in the EDD are scaled by a value of H$_0$=74.4 $\pm$ 3.0 km~s$^{-1}$Mpc$^{-1}$, the Mould flow parameters were also set to match the same cosmological parameters (H$_0$=74.4 $\pm$ 5.0 km~s$^{-1}$Mpc$^{-1}$, with $\Omega_m$=0.27 and $\Omega_\Lambda$=0.73). Table \ref{tab_props_with_dists}  lists the adopted distances and their associated uncertainties. The distance uncertainties were taken as the dominant contributor of uncertainty in the \mhi\ error budget. 
\section{Non-detections for the SFMS relation}
\begin{figure}
\centering
\includegraphics[scale=0.37]{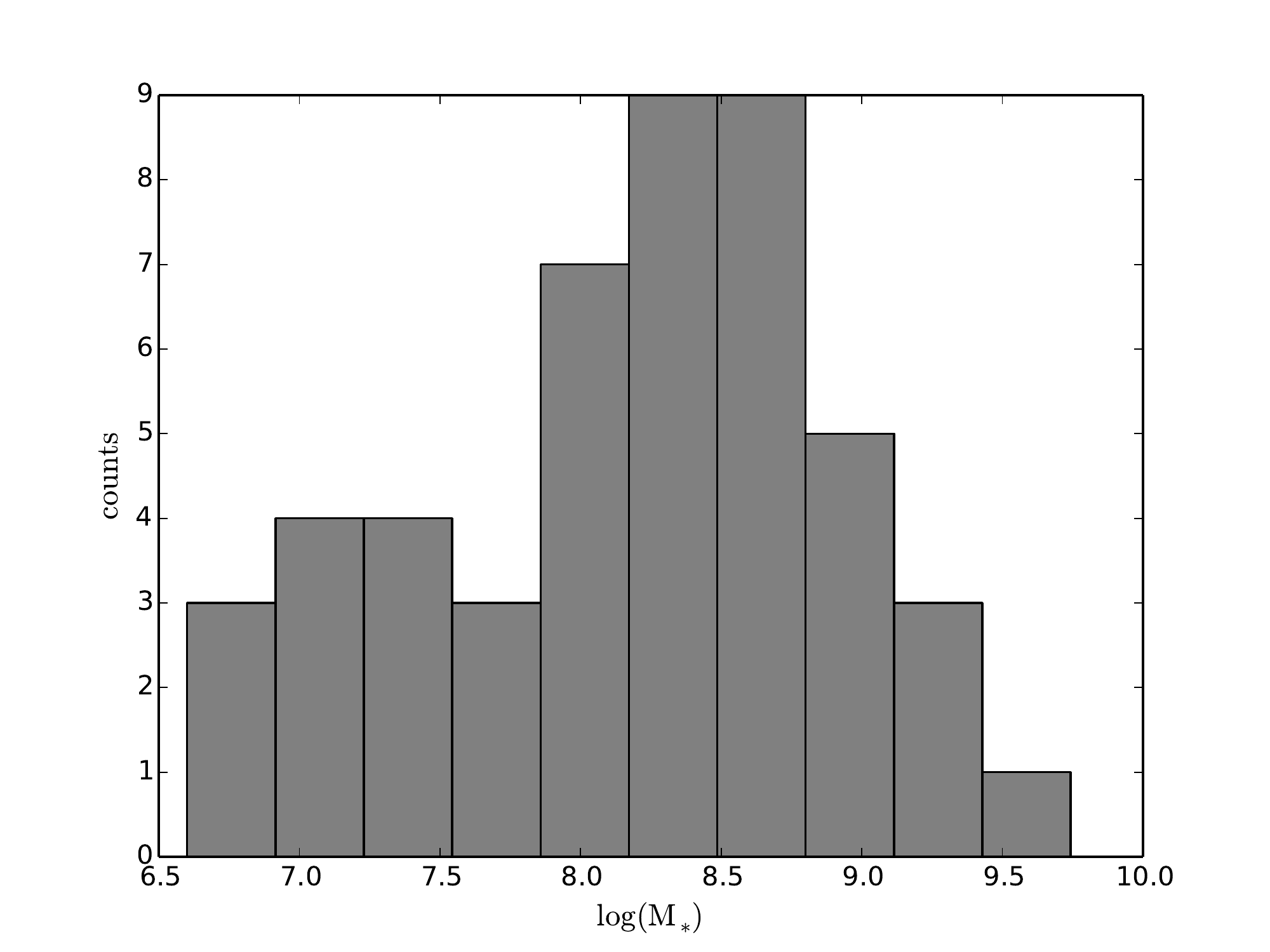}
\caption[Galaxies not detected in W3]{A histogram of galaxies in the main sample that were not detected in W3 and are thus not included in the analysis of star formation properties. These galaxies also did not have detections in the W4 band. The majority of them are dwarf galaxies and relatively dust-free, with very low MIR emission that lies below the detection threshold.}
\label{undetected}
\end{figure}

\begin{figure}
\centering
\includegraphics[scale=0.4]{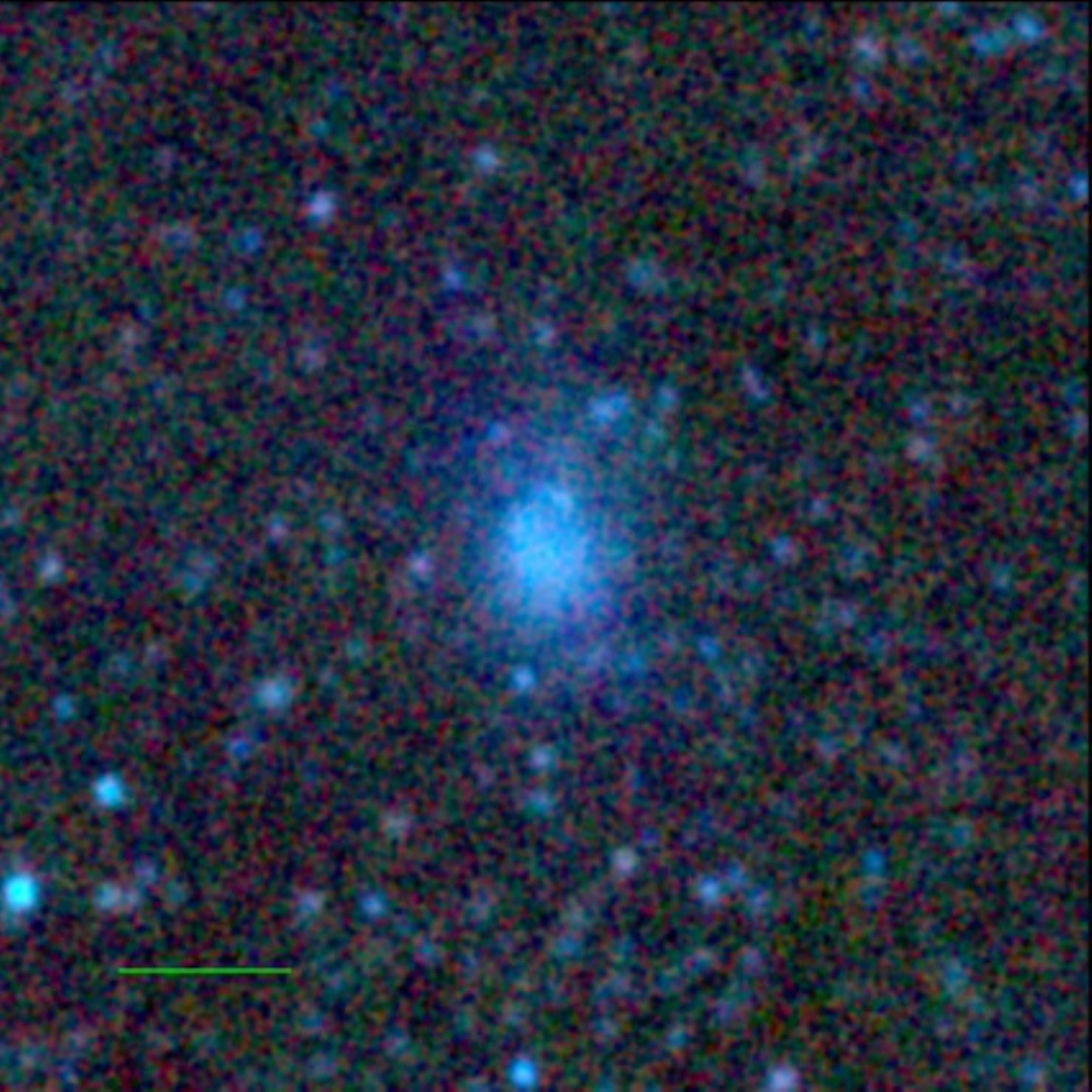}
\caption[u7125]{From our original sample of 228 galaxies, 48 (mostly dwarf) galaxies, were undetected in W3. Shown here is a WISE three-color (3.4 $\mmu$, 4.6 $\mmu$ and 12 $\mmu$) image of UGC7125, a late type dwarf galaxy. The green scale bar in the lower left shows an angular scale of 1 arcmin (1.5 kpc).}
\label{u7125}
\end{figure}

The most dominant flux source in the WISE 11.6 $\mmu$ band is continuum from warm dust grains which are more prevalent in the ISM of star forming spirals than in dwarf galaxies and dust-free ellipticals \citep{2017ApJ...850...68C}. As a result, 48 galaxies (mostly dwarfs) in our original sample were below the detection threshold of the W3 band. In Sections \ref{section:sfms} and \ref{section:sfrmhi}, we used a sub-sample of galaxies from our main sample which had detections in the 11.6 $\mmu$ band. The undetected galaxies were mostly low mass dwarfs and dust free spheroids. Figure \ref{undetected} below is a histogram of the stellar masses of the undetected galaxies, while Figure \ref{u7125} is an example galaxy detected in W1 but not in W3.
\newpage
\clearpage
%
\section{Global profile line widths}
Global profiles were obtained using the \textsc{MIRIAD} \textit{mbspect} task. The data points in the profiles were interpolated, and the 50\% and 20\% levels of the peak flux were located on either side of the global profile. The separation between the two velocities on either side of the profile at 50\% peak flux level was recorded as the $\mathrm{W_{50}}$ linewidth, and likewise for the $\mathrm{W_{20}}$ linewidth. The linewidths were corrected for instrumental broadening using the method of \citet{2001A&A...370..765V};
\begin{equation}
\begin{array}{r@{}l}
\mathrm{W_{20}^\prime}\ =\ \mathrm{W_{20} - 35.8\left[\sqrt{1\ +\ \left(\frac{R}{23.5}\right)^2}\ -\ 1\right]} \\
\\
\mathrm{W_{50}^\prime}\ =\ \mathrm{W_{50} - 23.5\left[\sqrt{1\ +\ \left(\frac{R}{23.5}\right)^2}\ -\ 1\right]},
\end{array}
\end{equation}
where R is the instrumental resolution in units of km/s, $\mathrm{W_{20}}$ and $\mathrm{W_{50}}$ are the observed linewidths, $\mathrm{W_{20}^\prime}$ and $\mathrm{W_{50}^\prime}$ are the corrected linewidths.
\section{Tables}
Tables \ref{tab_samp_A}, \ref{tab_props_with_dists}, \ref{tab_wise_phot} and \ref{tab_ks_samp} show the \HI\ observational parameters, the derived \HI\ global properties, WISE photometric parameters and derived global infrared properties of the sample respectively.
\begin{center}
\begin{table*}
\caption[Observational parameters of \HI\ data]{\textbf{Observational parameters of \HI\ data from the WHISP survey.}\\ Note: Col (2-3) - RA(J2000) and Dec(J2000), Cols (4-6) - Beam resolution (major and minor axis) and position angle. This study makes use of the WHISP data at resolution ~30" x ~30". Col (7) - Average noise in cubes before primary beam correction. Col (8) - Number of velocity channels, dependent on the WHISP flux density category($\mathrm{<100mJy, 100mJy-300mJy, >300mJy}$). Col (9) - Velocity resolution of the cube. (\textit{This table is available in it's entirety in the online journal.})}
\begin{tabular}{cS[table-format=3.4]S[table-format=2.4]S[table-format=2.2]S[table-format=2.2]S[table-format=2.0]S[table-format=1.3]cS[table-format=2.1]}
\hline \hline
Name & $\mathrm{R.A.}$ & {Dec} & $\mathrm{B_{maj}}$ & $\mathrm{B_{min}}$ & BPA & {Noise} & Channels & $\mathrm{\Delta V}$ \\
$\ $ & {deg} & {deg} & {$''$} & {$''$} & {deg} & $\mathrm{mJy\ beam^{-1}}$ & $\ $ & $\mathrm{km\ s^{-1}}$\\
(1) & {(2)} & {(3)} & {(4)} & {(5)} & {(6)} & {(7)} & (8) & {(9)}\\
\hline
UGC89 & 2.4725 & 25.9238 & 35.91 & 23.91 & 0 & 0.72 & 127 & 8.4 \\
UGC192 & 5.0723 & 59.3038 & 30.81 & 30.48 & 90 & 7.14 & 127 & 2.1 \\
UGC232 & 6.1612 & 33.2562 & 40.66 & 30.93 & 0 & 0.88 & 127 & 8.4 \\
UGC485 & 11.7845 & 30.3409 & 28.48 & 24.01 & 0 & 1.75 & 63 & 16.8 \\
UGC528 & 13.018 & 47.5505 & 30.0 & 30.0 & 0 & 4.26 & 127 & 4.1 \\
UGC624 & 15.1517 & 30.669 & 43.84 & 32.93 & 0 & 0.63 & 127 & 8.4 \\
UGC625 & 15.2309 & 47.682 & 28.33 & 28.32 & 0 & 1.96 & 63 & 16.6 \\
UGC655 & 16.0052 & 41.8429 & 31.87 & 26.08 & 0 & 4.15 & 127 & 4.1 \\
UGC690 & 16.8865 & 39.4001 & 29.98 & 25.17 & 0 & 1.79 & 63 & 16.8 \\
UGC731 & 17.6833 & 49.6022 & 30.0 & 30.0 & 0 & 4.3 & 127 & 4.1 \\
\hline
\end{tabular}
\label{tab_samp_A}
\end{table*}
\end{center}
\begin{center}
\begin{table*}
\caption[\HI\ global properties]{\textbf{\HI\ global properties.}\\ Note: Col (2) - Systemic velocity. Col (3-4) - Global profile linewidths at the 20$\%$ and 50$\%$ levels. Col (5) - \HI\ radius measured at \sighi\ = 1$\mathrm{M_\odot pc^{-2}}$. Col (6-7) - Integrated \HI\ flux and total \HI\ mass. Col (8) - Relative log uncertainty on mass. Col (9-10) - Distance in Mpc and associated uncertainty. Col (11) - The literature source of the distances, with NED for Mould flow model estimates from the NED database and CF2 for Cosmic flow database. (\textit{This table is available in it's entirety in the online journal.})}
\begin{tabular}{cS[table-format=4.1]S[table-format=3.1]S[table-format=3.1]cS[table-format=2.1]S[table-format=2.2]S[table-format=2.2]S[table-format=2.1]S[table-format=2.1]c}
\hline \hline
Name & $\mathrm{V_{sys}}$ & $\mathrm{W_{20}}$ & $\mathrm{W_{50}}$ & $\mathrm{R_{\HI}}$ & $\mathrm{\int{S dv}}$ & $\mathrm{M_{\HI}}$ & $\mathrm{e(M_{\HI})}$ & {D} & {e(D)} & Ref \\
$\ $ & $\mathrm{km\ s^{-1}}$ & $\mathrm{km\ s^{-1}}$ & $\mathrm{km\ s^{-1}}$ & $''$ & $\mathrm{Jy\ km\ s^{-1}}$ & {$\mathrm{\log(M_\odot)}$} & {Mpc} & {Mpc} & \\
(1) & {(2)} & {(3)} & {(4)} & (5) & {(6)} & {(7)} & {(8)} & {(9)} & {(10)} & {(11)}\\
\hline
UGC89 & 4509.8 & 425.2 & 371.9 & 82 & 6.8 & 9.66 & 0.17 & 53.7 & 10.7 & CF2 \\
UGC192 & -366.4 & 69.7 & 52.1 & 403 & 241.0 & 7.49 & 0.05 & 0.7 & 0.0 & CF2 \\
UGC232 & 4779.1 & 272.8 & 252.4 & 100 & 8.3 & 9.91 & 0.06 & 64.2 & 4.5 & NED \\
UGC485 & 5167.4 & 370.0 & 349.5 & 105 & 14.8 & 10.22 & 0.06 & 69.2 & 4.8 & NED \\
UGC528 & 661.7 & 129.2 & 82.2 & 100 & 13.7 & 8.6 & 0.06 & 11.1 & 0.8 & NED \\
UGC624 & 4709.6 & 543.4 & 509.1 & 121 & 21.8 & 10.5 & 0.17 & 78.3 & 15.7 & CF2 \\
UGC625 & 2630.7 & 348.0 & 328.6 & 174 & 37.3 & 9.93 & 0.17 & 31.2 & 6.2 & CF2 \\
UGC655 & 801.3 & 127.3 & 115.5 & 155 & 20.1 & 8.92 & 0.06 & 13.2 & 0.9 & NED \\
UGC690 & 5788.5 & 315.3 & 297.7 & 69 & 3.6 & 9.71 & 0.06 & 77.6 & 5.4 & NED \\
UGC731 & 618.7 & 143.4 & 129.6 & 188 & 44.4 & 9.12 & 0.06 & 11.3 & 0.8 & NED \\
\hline
\end{tabular}
\label{tab_props_with_dists}
\end{table*}
\end{center}

\begin{center}
\begin{table*}
\caption[Photometry of IR data]{\textbf{Photometric parameters of Infrared data from WISE.}\\ Note: Cols (2-3) - Axial ratio and position angle of the stellar disk, measured in the W1 band. Col (4) - Semi major axis of the 1$\sigma$ isophote in W1. Cols (5-10) W1, W3 and W4 Vega magnitudes, with corresponding magnitude errors, which were used to calculate the signal to noise ratio in the respective bands. The SNR in W3 was used to find a detection threshold for the sub-sample of 180 galaxies described in Section \ref{thesample}. Cols (11-12) - W1-W2 and W2-W3 color indices. Col (13) - Morphologies. (\textit{This table is available in it's entirety in the online journal.})}
\begin{tabular}{lS[table-format=1.2]S[table-format=3.1]S[table-format=3.2]S[table-format=2.2]S[table-format=2.2]S[table-format=2.2]S[table-format=2.2]S[table-format=2.2]S[table-format=2.2]S[table-format=2.2]S[table-format=2.2]l}
\hline \hline
Name & $\mathrm{b/a}$ & $\mathrm{pa}$ & $\mathrm{R1_{iso}}$ & $\mathrm{mag_1}$ & $\mathrm{\Delta mag_1}$  & $\mathrm{mag_3}$ & $\mathrm{\Delta mag_3}$  & $\mathrm{mag_4}$ & $\mathrm{\Delta mag_4}$  & $\mathrm{W1W2}$ & $\mathrm{W2W3}$ & Morph \\
{$\ $} & {$\ $} & {deg} & {$''$} & {mag} & {mag} & {mag} & {mag} & {mag} & {mag} & {mag} & {mag} & $\ $ \\
(1) & {(2)} & {(3)} & {(4)} & {(5)} & {(6)} & {(7)} & {(8)} & {(9)} & {(10)} & {(11)} & {(12)} & (13) \\
\hline
UGC89 & 0.65 & 169.3 & 100.03 & 8.57 & 0.01 & 4.81 & 0.01 & 2.37 & 0.01 & 0.13 & 3.63 & Sa \\
UGC192 & 0.87 & 125.8 & 440.03 & 5.11 & 0.01 & 2.32 & 0.01 & -0.32 & 0.01 & 0.08 & 2.71 & Im \\
UGC232 & 0.62 & 10.3 & 58.55 & 10.48 & 0.01 & 7.08 & 0.01 & 4.34 & 0.01 & 0.1 & 3.34 & Sa \\
UGC485 & 0.23 & 179.3 & 84.1 & 10.51 & 0.01 & 7.57 & 0.02 & 6.16 & 0.06 & 0.07 & 2.88 & Scd \\
UGC528 & 0.97 & 173.2 & 122.95 & 7.44 & 0.01 & 3.42 & 0.01 & 1.41 & 0.01 & 0.18 & 3.84 & Sb \\
UGC624 & 0.52 & 109.2 & 109.12 & 8.94 & 0.01 & 5.43 & 0.01 & 3.78 & 0.01 & 0.1 & 3.44 & Sab \\
UGC625 & 0.33 & 152.3 & 134.55 & 9.11 & 0.01 & 5.5 & 0.01 & 3.75 & 0.02 & 0.12 & 3.48 & Sbc \\
UGC655 & 0.82 & 130.4 & 58.62 & 11.46 & 0.01 & 10.99 & 0.09 & 10.12 & 0.78 & 0.01 & 0.57 & Sm \\
UGC690 & 0.72 & 93.9 & 69.02 & 10.22 & 0.01 & 7.03 & 0.01 & 5.4 & 0.03 & 0.07 & 3.15 & Scd \\
UGC731 & 0.56 & 68.5 & 88.97 & 11.61 & 0.01 & 13.4 & 0.23 & 11.63 & 1.54 & 0.12 & nan & Im \\
\hline
\end{tabular}
\label{tab_wise_phot}
\end{table*}
\end{center}
%

\begin{center}
\begin{table*}
\caption[Derived IR properties]{\textbf{Derived properties of the WISE data} \citep{2019arXiv191011793J}.\\ Note: Cols( 2-4) - The signal to noise ratios in the W1, W3 and W4 bands.  Cols (5-7) - Fluxes in the W1, W3 and W4. The W1 fluxes were not corrected for non-stellar sources of radiation, hence the derived stellar masses are to be taken as upper boundaries (see Section \ref{stellarmasssection}). Cols(8-9) - Stellar masses and associated (relative logarithmic) errors. Cols (10-13) - Star formation rates from the respective bands and corresponding errors. (\textit{This table is available in it's entirety in the online journal.})}
\begin{tabular}{cS[table-format=2.1]S[table-format=2.1]S[table-format=2.1]S[table-format=4.1]S[table-format=4.1]S[table-format=4.1]S[table-format=2.1]S[table-format=2.1]S[table-format=2.2]S[table-format=2.2]S[table-format=2.2]S[table-format=2.2]}
\hline \hline
Name &$\mathrm{SNR_{W1}}$ &$\mathrm{SNR_{W3}}$ & $\mathrm{SNR_{W4}}$ & {W1} & {W3} & {W4} & $\mathrm{\log (M_*)}$ & $\mathrm{e(M_*)}$  & $\mathrm{\log(SFR_{W3})}$ & $\mathrm{e(SFR_{W3})}$  & $\mathrm{\log(SFR_{W4})}$ & $\mathrm{e(SFR_{W4})}$ \\
$\ $& $\ $ & $\ $ & $\ $ & {mJy} & {mJy} & {mJy} & $\mathrm{M_\odot}$ & $\mathrm{M_\odot}$ & $\mathrm{M_\odot \ yr^{-1}}$ & $\mathrm{M_\odot \ yr^{-1}}$ & $\mathrm{M_\odot \ yr^{-1}}$ & $\mathrm{M_\odot \ yr^{-1}}$ \\
(1) & {(2)} & {(3)} & {(4)} & {(5)} & {(6)} & {(7)} & {(8)} & {(9)} & {(10)} & {(11)} & {(12)} & {(13)} \\
\hline
UGC89 & 98.7 & 90.5 & 90.5 & 110.4 & 327.7 & 904.1 & 10.78 & 0.1 & 1.03 & 0.15 & 1.17 & 0.17 \\
UGC192 & 98.7 & 90.5 & 90.5 & 2790.7 & 3110.0 & 10600.7 & 8.6 & 0.1 & -1.4 & 0.15 & -1.26 & 0.17 \\
UGC232 & 90.5 & 83.5 & 83.5 & 18.7 & 40.8 & 149.9 & 10.26 & 0.1 & 0.37 & 0.15 & 0.6 & 0.17 \\
UGC485 & 90.5 & 67.8 & 67.8 & 18.4 & 25.9 & 32.5 & 10.41 & 0.1 & 0.25 & 0.15 & 0.05 & 0.17 \\
UGC528 & 98.7 & 98.7 & 98.7 & 320.3 & 1202.3 & 2128.3 & 9.77 & 0.1 & 0.32 & 0.15 & 0.26 & 0.17 \\
UGC624 & 98.7 & 90.5 & 90.5 & 76.6 & 186.6 & 261.9 & 11.05 & 0.1 & 1.1 & 0.15 & 0.97 & 0.17 \\
UGC625 & 98.7 & 90.5 & 90.5 & 67.8 & 174.3 & 256.6 & 10.13 & 0.1 & 0.37 & 0.15 & 0.24 & 0.17 \\
UGC690 & 90.5 & 77.5 & 77.5 & 23.7 & 41.7 & 58.6 & 10.62 & 0.1 & 0.52 & 0.15 & 0.37 & 0.17 \\
\hline
\end{tabular}
\label{tab_ks_samp}
\end{table*}
\end{center}
%
\bsp	
\label{lastpage}
\typeout{get arXiv to do 4 passes: Label(s) may have changed. Rerun}
\end{document}